\documentclass[a4,11pt]{article}

\usepackage{harvard}
\usepackage[font={bf},justification=centering,labelsep=newline, position=top, belowskip=0.5mm]{caption}

\usepackage{amssymb}
\usepackage {amsmath}
\usepackage{amsthm}
\usepackage{graphicx}
\usepackage{url}
\usepackage{lscape}
\usepackage{multirow}

\usepackage{endnote} 

\newcommand \be{\begin{equation}}
\newcommand \bea{\begin{eqnarray}}
\newcommand \ee{\end{equation}}
\newcommand \eea{\end{eqnarray}}

\newcommand{\E}{{\rm E}}
\newcommand{\Var}{{\rm Var}~}
\newcommand{\Cov}{{\rm Cov}~}
\renewcommand{\(}{\left(}
\renewcommand{\)}{\right)}
\renewcommand{\[}{\left[}
\renewcommand{\]}{\right]}
\renewcommand{\epsilon}{\varepsilon}

\theoremstyle{plain}

\newtheorem{proposition}{Proposition}

\theoremstyle{definition}

\begin{document}

\title{A two-Factor Asset Pricing Model \\and the Fat Tail Distribution of Firm Sizes
\thanks{The authors acknowledge helpful discussions and 
exchanges with M.~Avellaneda, M.~Brennan, X.~Gabaix, M.~Grinblatt, M.~Meerschaert, V.~Pisarenko, R.~Roll, D.~Zajdenweber, W.~Ziemba and the seminar participants at New York University. All remaining errors are ours.}}

\thispagestyle{empty}

\author{Y. Malevergne$^{1,2}$ and D. Sornette$^1$\\
\\
$^1$ ETH Zurich -- Department Management, Technology and Economics, Switzerland\\
$^2$ EM-Lyon Business School -- Department Economics, Finance and Control, France\\
e-mails: ymalevergne@ethz.ch and dsornette@ethz.ch
}

\date{}
\maketitle

\abstract{
We derive a theoretical two-factor model which has empirically a similar explanatory power as the Fama-French three-factor model. In addition to the usual market risk, our model accounts for a diversification risk, proxied by the equally-weighted portfolio, and which results from an ``internal consistency factor'' appearing for arbitrary large economies, as a consequence of the concentration of the market portfolio when the distribution of the capitalization of firms is sufficiently heavy-tailed as in real economies. Our model rationalizes the superior performance of the Fama and French three-factor model in explaining the cross section of stock returns: the size factor constitutes an alternative proxy of the diversification factor while the book-to-market effect is related to the increasing sensitivity of value stocks to this factor.

}

\section*{Introduction}
In the standard equilibrium and/or arbitrage pricing framework, the
value of any asset is uniquely specified from the belief that only the
systematic risks need to be remunerated by the market. This is the
conclusion of the CAPM
\cite{Treynor1961,Treynor1962,Sharpe1964,Lintner1965,Mossin1966} and of
the APT \cite{Ross1976,RollandRoss1980,RollandRoss1984,Roll1994}. Here,
we show that, even for arbitrary large economies when the distribution
of the capitalization of firms is sufficiently heavy-tailed as is the
case of real economies,
there may exist a new source of significant systematic risk, which has
been totally neglected up to now but must be priced by the
market. This new source of risk can readily explain several asset
pricing anomalies on the sole basis of the internal-consistency of the market model.
 
This result is based on two ingredients. The first one
is the tautological internal consistency condition that the market portfolio, 
and any other factor that can be replicated by a portfolio of assets traded on the market, 
is constituted -- by construction -- of the assets whose returns it is supposed to explain.
This internal consistency condition leads mechanically to
correlations between the return residuals, as already stressed by
\citeasnoun{Fama1973} and \citeasnoun[footnote 13]{Sharpe1990} when the
return on the market portfolio is considered as the only explaining
factor, or by \citeasnoun{Chamberlain1983} in the case where there exists
several linearly independent portfolios that contain only ``factor''
variance and are therefore optimal for any risk-averse investor. These
correlations are equivalent to the existence of at least one internal
consistency factor (uncorrelated with the market and
the other explanatory factors), which is a function of the weights of the 
market portfolio and of the portfolios replicating the other factors.
The impact of this new factor is usually neglected away
on the basis of the law of large numbers applied to well-diversified
portfolios.

Actually, when the distribution of the weights of the portfolios
replicating the explaining factors -- the distribution of the
capitalization of firms in the case of the market portfolio, for
instance -- is sufficiently heavy-tailed, the law of large numbers, which is at the origin of
the diminishing contribution of the
residual risks to the total risk of ``well-diversified portfolios''
\cite{Ross1976,Huberman1982}, breaks down. Intuitively, whatever the
size of the economy, the largest firms contribute idiosyncratic risks
that can not be diversified. In this case, the generalized central limit
theorem \cite{GnedenkoandKolmogorov1954} shows that the impact of an
internal consistency factor does not vanish even for infinite
economies\footnote{In a different context, \citeasnoun{Gabaix2005} has
proposed that the same kind of argument can explain that idiosyncratic
firm-level fluctuations are responsible for an important part of
aggregate shocks, and therefore provide a microfoundation for aggregate
productivity shocks. Indeed, as in the present article, it is suggested
that the traditional argument according to which individual firm shocks
average out in aggregate breaks down if the distribution of firm sizes
is fat-tailed, as documented empirically.}. This
may be the origin of a significant amount of risk
for portfolios that would have been otherwise assumed ``well-diversified'' in its
absence. As a consequence, when writing down for instance the APT, 
an additional explaining factor must be accounted for.

This result must be contrasted with the many seminal papers deriving the
APT and providing pricing bounds for finite economies. Indeed, following
for instance \citeasnoun{Dybvig83} or \citeasnoun{GT83}, among others,
the residual risk of well-diversified portfolios resulting from the
finiteness of the economy should be priced but the pricing error
relative to a pure factor model disappeared in the limit of a large
economy, as a full diversification of the non-systematic risk is
achieved. In contrast, we find that the lack of diversification persists
even when the number of traded assets is infinite. Beside, the
generalization of \citeasnoun{Ross1976}'s results provided by
\citeasnoun{Chamberlain1983} breaks down as a result of this lack of
diversification. Indeed, \citeasnoun{Chamberlain1983}'s results
explicitly require that the risk of any sequence of portfolios bearing
only residual risks converges to zero if the portfolios are
well-diversified. Similarly, one cannot apply anymore 
\citeasnoun{Connor1982}'s result that 
the APT pricing equation holds exactly if each asset
has an infinitesimal weight in the economy. Indeed, in
economies with a heavy-tailed distribution of firm sizes, the largest
company has a size of the same order as the total size of all the
companies\footnote{This simply results from the large deviation theorem on
heavy-tailed distribution according to which, given $N$ iid random
variables $S_1$, \dots, $S_N$ with a fat tailed distribution, we have
\cite{Embrechtsetal1997}
$$
\lim_{x \rightarrow \infty} \frac{ \Pr \[\max (S_1, \dots S_N) \ge x\]
}{\Pr \[S_1 + \cdots + S_N \ge x\]} = 1.
$$}.
These different remarks are in fact intimately entangled as will become clear
in the sequel of this article. We stress that our
results are driven by the fat-tailed nature of the distribution of
the weights of the portfolios replicating the
factors (when replication is possible), as occurs for the market portfolio when the
distribution of firm sizes is heavy-tailed. Our results do not rely on
any other distributional assumption concerning the explanatory factors
or the disturbance terms. For simplification, we will assume that both
the factors and the disturbance terms have finite variance, but it is
simply for the convenience of the exposition of our results. They could
easily be generalized to the case where factors and disturbance terms do not admit a finite second
moment on the basis of the result established by \citeasnoun{Wang1988},
for instance.

The introduction of our new ``internal consistency factor,'' which
basically accounts for the lack of diversification of  the market
portfolio, allows us to provide a theoretical explanation of several
well-known pricing anomalies. In particular, the relevance of the two
effects studied by Fama and
French~\citeyear{FamaandFrench1992,FamaandFrench1993,FamaandFrench1995},
namely the small-firm effect (first documented by \citeasnoun{Banz}) and the
book-to-market ratio, can
be understood from and rationalized within the theoretical framework of the ATP
when the ``internal consistency factor'', and its associated
diversification premium, is accounted for. Thus, our model bridges the
gap between Fama and French phenomenological model and the arbitrage
pricing theory. More precisely, our model provides an
understanding of the superior performance of Fama and French's three-factor model
in explaining the cross section of stock returns.  Indeed, the new
internal consistency factor provides a rationalization of the size
factor as a proxy of the internal consistency factor. Besides, consistent
with the fact that high book-to-market stocks have significantly lower
beta's with respect to the market portfolio compared with low
book-to-market stocks \cite{Bernardoetal2005}, the book-to-market effect
also emerges naturally from our formalism. In the context of the on-going
debate \cite{Lako,Daniel} on the interpretation of the two 
empirical effects analyzed by \citeasnoun{FamaandFrench1993}, we provide an 
explanation with solid economic underpinning.

The article is organized as follows. In the next section, we synthesize
the available empirical evidence on the fat-tailed nature of the
distribution of firm sizes and their consequence on the lack of
diversification of the market portfolio. Then, in section 2, we make
clear the consequences of the internal consistency condition mentioned
above; due to the internal consistency condition, we show that the
disturbance terms must obey a condition which determines their
correlation. Next we presents our main results on the asymptotic
behavior of the variance of well-diversified portfolios: we show that,
together with the market risk, there is an additional source of
systematic risk resulting from the internal consistency condition. This
additional risk may be of the same order as the market risk even for
infinite economies when the distribution of the capitalization of
companies is sufficiently heavy-tailed. Section 3 confirms, by use of
numerical simulations, the relevance of the concentration effect for
markets with a realistic number of traded assets. Then, it  discusses
the consequences for the arbitrage pricing of financial assets,
providing an
expression that accounts for the premium required by the investors to
bear this systematic ``internal consistency'' risk and we propose
proxies for the empirical assessment of this risk premium. It allows us
to provide theoretical economic explanations of some of the empirical
factors reported in the literature while an empirical analysis
shows that, on the basis of only two factors (the market portfolio
and the equally-weighted portfolio), our model is at least
as successful as the Fama and French three-factor model over the period
Jan. 1927 to Dec. 2005 for the US market data available on Professor
French's
website\footnote{\url{http://mba.tuck.dartmouth.edu/pages/faculty/ken.
french/data_library.html}}. Section 4 summarizes our results and draw some conclusion.

\section{The distribution of firm sizes and the concentration of the market portfolio}
\label{sec1}

The study of the distribution of firm sizes benefits from a rich
history. \citeasnoun{Zipf1949} made an important early contribution by
establishing that US corporation assets approximately followed the law
$s(n) \sim 1/n$ (now referred to as the Zipf's law): when sizes are
ranked from the largest to the smallest, Zipf's law states that the firm
size $s(n)$ of the $n^{th}$ largest firm is inversely proportional to
its rank $n$. Inverting this relation, we have that the rank of the
$n^{th}$ largest firm is inversely proportional to its size $n \sim
1/s(n)$ which is nothing but the sample complementary cumulative
distribution of the Pareto law with a tail exponent $\mu=1$.

Zipf's law seems to be a robust property of business firms\footnote{
Other social entities, such as cities, share this property
\cite{Zipf1949,Gabaix1999a,Gabaix1999b,GabaixandIoannides2004}.}
\cite{IjriandSimon1977}. Indeed, several proxies for the size of
companies have been used which recover the same robust results that the
exponent $\mu$ is equal or close to $1$: assets, market capitalizations,
number of employees, profits, revenues, sales, value
added and so on \cite{Axtell2001,Axtell2006,Gabaixetal2003,Marsili2005,SimonandBonini1958}. 
Beside, \citeasnoun{RamsdenandKiss-Haypa2000} have analyzed the
distribution of firms by revenues in 20 countries in America, Asia and
Europe and report an exponent $\mu$ ranging from $0.44$ to $1.25$
with a median value equal to $0.85$.

Several models have attempted to provide explanations for the
distribution of firm sizes, in terms of the law of proportional effect
\cite{Gibrat1931,SimonandBonini1958}, of economies of scale and costs
reduction \cite{Bain1956,Robinson1942}, of the distribution of
managerial talents and efficient allocation of productivity factors
across managers \cite{Lucas1978}, or of the partition of the set of
workers \cite{Axtell2006}, among others. But, only recently, the
closeness of the exponent $\mu$ to the value $1$ has been justified from
a simple argument proposed by \citeasnoun{Gabaixetal2003}. They have
transposed the mechanism given for cities \cite[and the references
therein]{Gabaix1999a,GabaixandIoannides2004} to firm sizes and mutual
fund capitalizations: starting from the traditional argument based upon Gibrat's law of proportional effect,
whereby firm growth is treated as a random process and growth rates are
independent of firm size, a log-normal process modified with small
perturbations to ensure convergence to a non-degenerate steady-state
distribution yields a power law distribution. The value of its exponent
$\mu=1$ then results from the condition that the average normalized size of
firms stays constant in a stationary economy.

Consubstantial with the fat tailed character of the distribution of firm
sizes is the concentration of the market portfolio. Indeed, the market
portfolio, defined as the value-weighted portfolio of all the assets
traded on a given market suffers from an inherent lack of
diversification, resulting from the fat tail distribution of firm sizes,
in the sense that only a few dozen of companies account for a very large
part of the overall market capitalization. For instance,  the top ten
largest companies of the US market represents about one fifth to one
fourth of the US market capitalization.

More generally, given an economy of $N$ firms, whose sizes $S_i$,
$i=1,\dots, N$, follow a Pareto law with tail index $\mu$, the ratio of
the capitalization of the largest firm to the total market
capitalization
\be
R_N=\frac{\max S_i}{\sum_{i=1}^{N} S_i}
\ee
which is nothing but the weight of the largest company in the market
portfolio, behaves on average like
\bea
\E \[R_N\] &\longrightarrow& 0, \quad \text{if } \mu>1,\\
\E \[1/R_N\] &\longrightarrow& \frac{1}{1-\mu},\quad \text{if } \mu<1,
\eea
as the number of firms $N$ goes to infinity \cite{Betal87}.

This result means that when the distribution of firm sizes admits a
finite mean, the weight of the largest firm in the market portfolio goes
to zero, and so do the weights of any other firms, in the limit of a
large market. In terms of asset pricing, the fact that the weight of
each individual firm in the economy is infinitesimal ensures that the
APT pricing equation holds for each asset and not only on average
\cite{Connor1982}. In contrast, when the distribution of firm sizes
has no finite mean, the asymptotic weight of the largest firm in the
market portfolio does not vanish, illustrating the fact that for such
an economy, the market portfolio is not well diversified, all the more
so the smallest the tail index $\mu$. A practical consequence is then
that the APT pricing equation, if it holds, only holds on average, with possibly large
pricing errors for individual assets.

In order to get a closer look at the concentration of the market
portfolio, we focus on its Herfindahl index, which is perhaps the most
widely used measure of economic concentration \cite{Polakoff81,Lovett88},
\be
H_N = ||w_m||^2 = \sum_{i=1}^{N} w_{m,i}^2~,
\ee
where $w_{m,i}$ denotes the weight of asset $i$ in the market portfolio
whose composition is given by the $N$-dimensional vector $w_m$. 
The Herfindahl takes into account the relative size and distribution of
the firms traded on the market. It approaches zero when the market
consists of a large number of firms with comparable sizes. It
increases both as the number of firms in the market decreases and as the
disparity in size between those firms increases. Our use of the
Herfindahl index is not only guided by common practice but also by
its superior ability to provide meaningful information about the degree
of diversification of an unevenly distributed stock portfolio
\cite{WP93}. Following tradition, we say that a portfolio is {\em
well-diversified}, if its
Herfindahl index goes to zero when the number $N$ of firms traded in the
market goes to infinity. 

For illustration purpose, let us first concentrate on an economy where
the sizes, sorted in descending order, of the $N$ firms are
deterministically given by
\be
S_{i,N} = \( \frac{i}{N}\)^{-1/\mu}.
\ee
We have arbitrarily chosen the size of the smallest firm as equal to
one. Alternatively, one can think of $S_{i,N}$ as the size of the
$i^{th}$ largest firm relative to the size of the smallest one. 
With this simple model, the rank $i$ of the $i^{th}$ largest
company is directly proportional to its size taken to the power of minus
$\mu$, as it should in order for the distribution of sizes to obey a Pareto law
with a tail index equal to $\mu$. It is easy to check that
the weight of the largest firm in the market portfolio goes to zero, as
$N$ goes to infinity, when $\mu$ is larger than or equal to one while it
goes to some positive constant when $\mu$ is less than one. More
precisely, we have
\bea
w_{m,1} &\longrightarrow& 0, \quad \text{if } \mu \ge 1,\\
w_{m,1} &\longrightarrow& \frac{1}{\zeta\(1/\mu\)},\quad \text{if } \mu<1,
\eea
where $\zeta(\cdot)$ denotes the Riemann zeta function \cite[p. 807]{AS1972}. 

For the Herfindahl index, one gets
\be
H_N = 
\left\{
\begin{array}{lr}
\displaystyle \frac{1}{1-\frac{1}{(1-\mu)^2}} \cdot \frac{1}{N} + O \(N^{2/\mu-2} \),& \quad \mu >2,\\
\displaystyle \frac{\ln N + \gamma}{4N} + O \(N^{-3/2} \ln N \),& \quad \mu = 2,\\
\displaystyle \(\frac{1-\mu}{\mu}\)^2 \zeta(2/\mu) \cdot \frac{1}{N^{2-2/\mu}} + O\( N^{3(1/\mu -1)} \),& \quad 1 <\mu < 2,\\
\displaystyle \frac{\pi^2}{6} \frac{1}{\(\gamma + \ln N \)^2} + O\(N^{-1}(\gamma + \ln N )^{-2} \),& \quad \mu =1,\\
\displaystyle \frac{\zeta(2/\mu)}{\zeta(1/\mu)^2} + O\(N^{1-1/\mu} \),& \quad \mu < 1 .
\end{array}
\right.
\ee
In accordance with the behavior of the weight of the
largest firm, $H_N$ goes to zero when the index $\mu$ is larger than or
equal to one, while it goes to some positive constant otherwise. However,
the decay rate of $H_N$ toward zero becomes slower and slower as
$\mu$ approaches $1$ (from above). In practice, when the
number of traded firms is large -- but finite -- the concentration of
the market portfolio can remain significant even if $\mu$ is larger than
one, specifically when $\mu$ lies between one and two.

\begin{figure}
\begin{center}
\includegraphics[width=12cm]{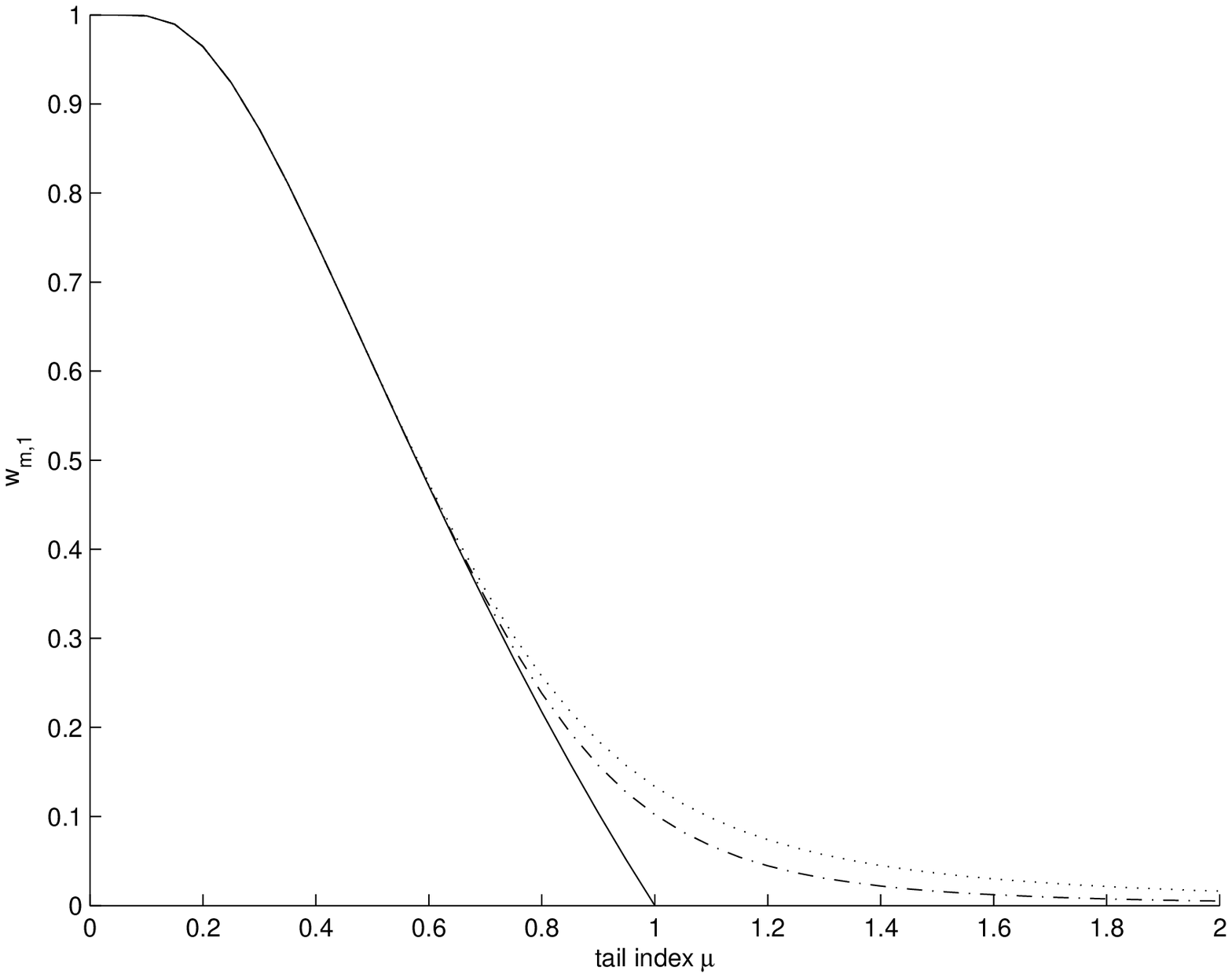}\\
\includegraphics[width=12cm]{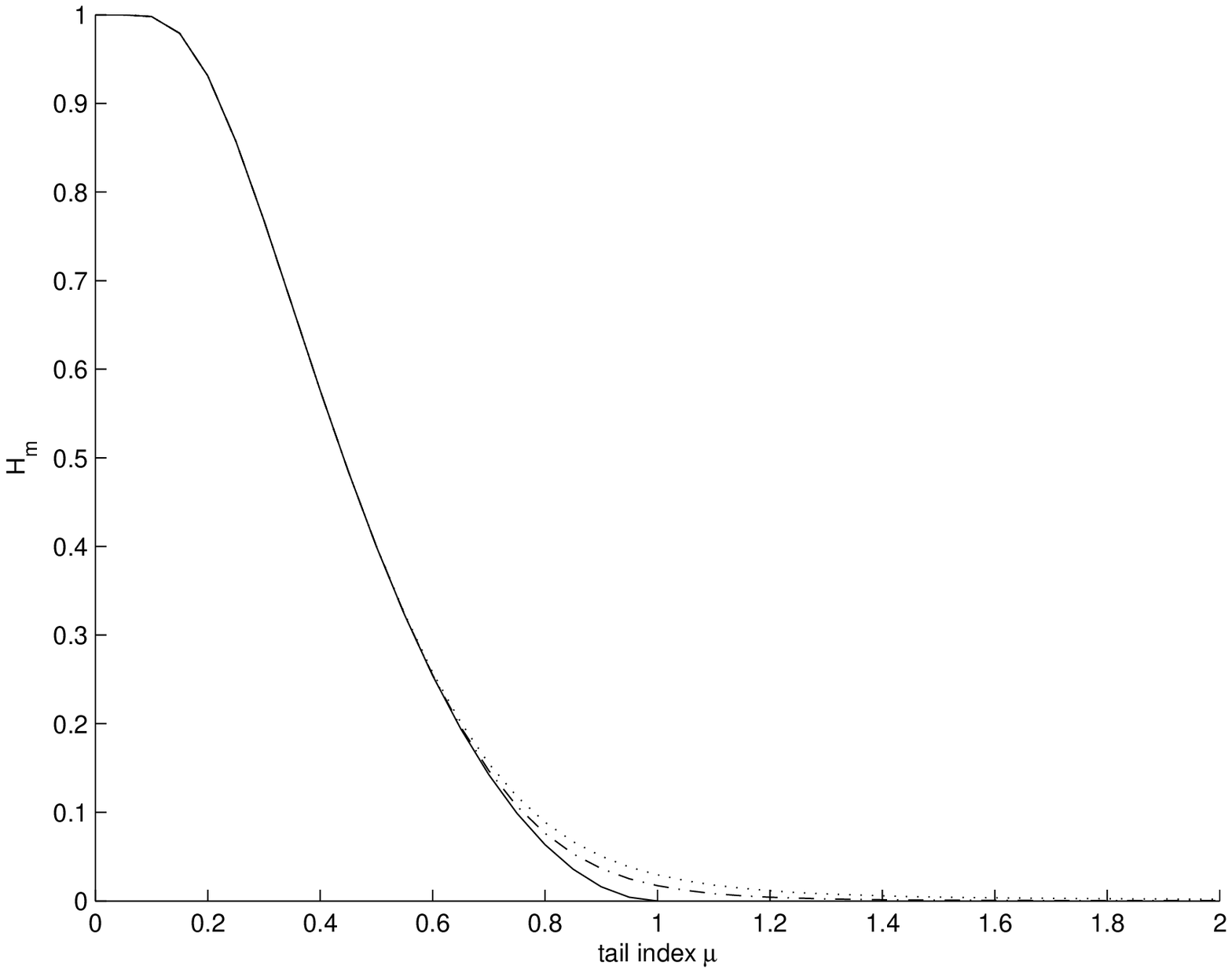}
\end{center}
\captionsetup{style=default,labelfont=bf, labelsep=period}
\caption{\label{FigWmHm} {\bf Concentration of the market portfolio}.
The upper panel shows the weight of the largest firms in the market
portfolio as a function of the tail index $\mu$ of the Pareto distribution of
firm sizes. The lower panel shows the Herfindahl index of the market
portfolio as a function of the tail index $\mu$ of the Pareto distribution of
firm sizes. In both cases, the continuous line provides the values in the
limit of an infinite economy while the dotted and dash-dotted curves
refers to the cases of an economy with one thousand and ten thousand
firms respectively. }
\end{figure}

In order to illustrate this situation, the upper panel of
figure~\ref{FigWmHm} depicts the value of the weight of the largest firm
in the market portfolio while the lower panel shows the Herfindahl index
as a function of $\mu$. The plain curves show the limit situation of an
infinite economy while the dotted and dash-dotted curves account for 
the effect of a finite economy: the dotted curve refers to the case where only one
thousand companies are traded while the dash-dotted curve corresponds to an
economy with ten thousand firms. Clearly, finite economy size effects cannot be
neglected for market sizes as found in the real economy.

\vspace{0.5cm}
\centerline{\it [Insert Figure~\ref{FigWmHm} about here]}
\vspace{0.5cm}

To be a little bit more general, we now consider the case where the firm
sizes are randomly drawn from a power law distribution of size. By
application of the generalized law of large numbers
\cite{Feller71,GnedenkoandKolmogorov1954,IL75} and using standard results on
the limit distribution of self-normalized sums \cite{Darling52,LMRS73},
we can state that
\begin{proposition}
\label{prop3a}
The asymptotic behavior of the concentration index $H_N$ is the following:
\begin{enumerate}
\item provided that $\E[S^2] < \infty$,
$$H_N = \frac{1}{N}
\frac{\E\[S^2\]}{\E\[S\]^2} + o_p(1/N),$$
\item provided that $S$ is regularly varying with
tail index $\mu=2$ and $s^{\mu}\cdot \Pr\[S>s\] \rightarrow c$ as $s \rightarrow \infty$, 
$$
H_N = 
\frac{c}{\E\[S\]^2}\frac{\ln N}{N} + o_p\( \frac{1}{N \ln N} \),
$$
\item provided that $S$ is regularly varying with tail index $\mu \in
(1,2)$ and $s^{\mu}\cdot \Pr\[S>s\] \rightarrow c$ as $s \rightarrow
\infty$,
$$
H_N = \[\frac{\pi c}{2 \Gamma \(\frac{\mu}{2}\) \sin \frac{\mu \pi}{4}}\]^{2/\mu}
\frac{1}{\E\[S\]^2}
\cdot\frac{1}{N^{2-2/\mu}}
\cdot \xi_N + o_p\(\frac{1}{N^{2-2/\mu}}\)~,
$$ 
where $\xi_N$ is a
sequence of positive random variables with stable limit law\footnote{The
stable law $S(\alpha,\beta)$ has characteristic function
$\psi_{\alpha,\beta}(s) = 
\begin{cases}
\exp \[ - |s|^{\alpha} + i s \beta  \tan \frac{\alpha \pi}{2} |s|^{\alpha - 1} \] & \alpha \neq 1,\\
\exp \[ - |s| - i s \beta  \frac{2}{\pi} \cdot \ln s\] & \alpha = 1,
\end{cases}$ with $\beta \in [-1,1]$.} $S(\mu/2,1)$,
\item provided that $S$ is regularly varying with tail index $\mu = 1$
and $s^{\mu}\cdot \Pr\[S>s\] \rightarrow c$ as $s \rightarrow \infty$,
$$
H_N
= \frac{\pi}{2 \cdot {\ln^2 N}} \cdot \xi_N +
O_p\(\frac{1}{\ln^3 N}\),
$$
where $\xi_N$ is a sequence of positive random variables with stable 
limit law $S(1/2,1)$,
\item provided that $S$ is regularly varying with tail index $\mu
\in(0,1)$ and $s^{\mu}\cdot \Pr\[S>s\] \rightarrow c$ as $s \rightarrow
\infty$,
$$
H_N =  \frac{4}{\pi^{1/\mu}}\[ \Gamma\(\frac{1+\mu}{2}\) \cos \frac{\pi
\mu}{4}\]^{2/\mu}
\cdot \frac{\xi_N}{\zeta_N^2},
$$
where $\xi_N$ and $\zeta_N$ are two sequences of strongly correlated\footnote{
\label{foot}
More precisely, the sequence of random vectors $(\xi_N,\zeta_N)'$ converges to an {\em operator}-stable law with stable marginal laws $S(\mu/2,1)$ and $S(\mu,1)$ respectively, and a spectral measure concentrated on arcs $\pm(x,x^2)$. The full characterization of the spectral measure is beyond the scope of this article (see \cite[Section 10.1]{MS2001} for details).}
positive random variables that converge in law to $S(\mu/2,1)$ and $S(\mu,1)$ respectively,
\item provided that $S$ is slowly varying\footnote{The random variable
$S$ is slowly varying if its distribution function $F$ satisfies
$\lim_{x \rightarrow \infty} \frac{1-F(tx)}{1-F(x)}=1$, for all $t>0$.
It corresponds to the limit case where $S$ is regularly varying with
$\mu \rightarrow 0$.},
$$
H_N \rightarrow 1, \quad  \text{a.s.}
$$
\end{enumerate}
\end{proposition}

As a consequence of the fourth statement of the proposition above, for
economies in which the distribution of firm sizes follows Zipf's law
($\mu=1$) the asymptotic behavior of the concentration index $H_N$ of
the market portfolio is given by
\be
H_N \simeq \frac{\pi}{2 \cdot \(\ln N\)^2} \cdot \xi_N, 
\ee
where $\xi_N$ is a sequence of positive random variables with stable
limit law $S(1/2,1)$, namely the L\'evy law with density
\be
\label{eq:LevyLaw}
f(x)=\frac{1}{\sqrt{2\pi}}\cdot x^{-3/2} {\rm e}^{-\frac{1}{2x}}, \quad x\ge 0.
\ee
This shows that, even if the concentration of the market portfolio goes to
zero in the limit of an infinite economy, it goes to zero extremely
slowly as the size $N$ of the economy diverges. Accounting for the fact
that the median value of the L\'{e}vy law (\ref{eq:LevyLaw}) is approximately equal to
$2.198$, a typical value of $H_N$ is $4-5\%$ for a market where $7000$
to $8000$ assets are traded\footnote{These figures are compatible with
the number of stocks currently listed on the Amex, the Nasdaq and the
Nyse.}, which is much higher than the concentration index of a
well-diversified portfolio -- typically the equally-weighted portfolio --
which should be of the order of $0.012-0.014\%$. Intuitively, $H_N
\simeq 4-5\%$ means that there are only about $1/H_n \simeq 20-25$ 
effective assets in a typical portfolio supposedly well-diversified on $7000-8000$
assets.

This simple illustrative example shows, roughly speaking, that the
market portfolio reflects the behavior of the 20 to 25 largest assets
traded on the market. In this context, one can wonder (i) how the market
portfolio alone could explain the expected return on any asset,
irrespective of its size, as predicted by the CAPM and (ii) if it is
actually optimal for a rational investor to put her money in this risky
portfolio alone, as proposed by the two-fund separation theorem. 
This suggests that the lack of diversification of the
market portfolio is responsible, to a large extent, for the failure of
the CAPM to explain the cross-section of stock returns. This failure
has been documented
in particular by Fama and French \citeyear{FamaandFrench1992,FamaandFrench1993},
who find basically no support for the CAPM's central result of a positive
relation between expected returns and the global market risk (quantified
by the beta parameter). This therefore raises the question of the
existence of a {\em concentration premium}.

Many authors have proposed alternative or additional factors in the
quest to cure the deficiencies of the CAPM and provide explanations for
the so-called {\em pricing anomalies}. Three main
classes of additional factors can be distinguished: macro-economic factors, 
firm-specific factors and behavioral factors.
\begin{itemize}
\item[] {\em Macro-economic factors.}
The positive or negative impact on stock prices of macro-economic factors such as interest rates \cite{CRR86}, exchange rates \cite{Harvey91,FH93}, real output \cite{CPS89,CRR86}, inflation and money supply \cite[1985]{Bodie76,Fama81,GR83,PR83}, aggregate consumption \cite[and references therein]{JW07}, oil prices \cite{CRR86,FH93,JK96}, labor income \cite{JW96,Reyfman97} and so on, has been underlined in many studies based on the APT
\cite{Ross1976,RollandRoss1984,Roll1994} or in the context of
equilibrium \cite{Burmeister86,Flannery02}.

\item[] {\em Firm-specific factors.}
The fact that industry sector groupings may be important in the study of
the return generating process has been stressed for a long time
\cite{King66,Alex86}. Similarly, the importance of market capitalization
(or small-firm effect) has been documented in the early eighties by
\citeasnoun{Banz} and \citeasnoun{Reinganum81} while
\citeasnoun{Stattman80} and \citeasnoun{RRL85} underlined the role of the
book-to-market ratio. If other ratios such as the earnings-to-price
ratio \cite{Basu77} and the dividend yield
\cite{Blume80,Rozeff84,Keim85} for instance, also predict
future returns, most of the attention has been drawn to the size and the
book-to-market effect during the past decade as a result of their
superior performance to explain the cross-section of stock returns
\cite[1993,1995,1996]{FamaandFrench1992}. Among various interpretation
of the explaining power of the size and the book-to-market ratio,
\citeasnoun{CV2004} and \citeasnoun{CPV2004}
have considered breaking the beta of a stock with the market portfolio
into two components, one reflecting news about the market's future cash
flows and one reflecting news about the market's discount rates in order
explain the size and value ``anomalies'' in stock returns.

\item[] {\em Behavioral factors.}
Two major issues have been considered. On the one hand,
\citeasnoun{Rubinstein1973} and \citeasnoun{KraussandLitzenberger1976}
have proposed to account for the departure of the distributions of returns
from normality and for the sensibility of the investors for the skewness
and kurtosis of the distribution of stock returns. The relevance of this
approach has been underlined by \citeasnoun{Lim1989} and
\citeasnoun{HarveyandSiddique2000} who have tested the role of the
asymmetry in the risk premium by accounting for the skewness of the
distribution of returns. Along the same line, many other extensions have
been presented such as the VaR-CAPM \cite{AlexanderandBaptista2002}, in
order to account more carefully for the risk perception of investors. On
the other hand, several studies have developed
phenomenological models capturing the reversal of long-term returns
\cite[1987]{Chan88,CLR92,DBT85} and the continuation of short-term trends
\cite{CJL96,JT1993,JT2001,Richards97}.
\end{itemize}

Most of these factors actually provide a significant improvement in explaining
the cross-section of asset returns. However they do not provide a
clear identification of the most prominent ones. Even if the Fama and
French three factor model is now widely recognized as the benchmark, the
reasons for its superiority in explaining the cross-section of asset
returns are still debated. It is in this context that we propose to focus on the
consequences of the undisputable fact that the market portfolio is
highly concentrated on a small number of very large companies and
therefore can obviously not account for the behavior of the smallest
ones. As we are going to demonstrate, this will allow us to rationalize
the size effect, in relation with what we propose to call a
``diversification factor,'' which, to some extent, also justify the
relevance of the book-to-market factor.

\section{Internal consistency conditions of factor models and their
consequences on diversification}
\label{sec:capm}
\label{sec:Diversification}

Under the assumption that the return on the market portfolio is a factor
explaining the return on individual assets, our demonstration is based on two ingredients. 
\begin{itemize}
\item The internal consistency condition states that the market portfolio is
made of the assets whose returns it is supposed to explain.
As a consequence, there are correlations between the disturbance terms.
\item The lack of diversification of the market portfolio
(associated with the fat tail distribution of firm sizes) make these
correlations non-negligible, giving birth to an additional factor which
significantly contributes to the asymptotic variance of {\it a priori}
well-diversified portfolios.
\end{itemize}

\subsection{The factor model}

Consider an economy with $N$ firms whose returns on stock prices
are determined according to the following equation
\be
\label{eq:capm}
\vec r = \vec \alpha + \vec \beta_m \cdot \[r_m - \E\[r_m\]\] + B \vec \phi + \vec \epsilon,
\ee
where
\begin{itemize}
\item $\vec r$ is the random $N \times 1$ vector of asset returns;
\item $\vec \alpha = \E \[\vec r\]$ is the $N \times 1$ vector of asset
return mean values. We do not make any assumption neither on the {\it
ex-ante} mean-variance efficiency of the market portfolio, nor on the
absence of arbitrage opportunity, so that $\vec \alpha$ is not, {\it a
priori}, specified;
\item $r_m$ is the random return on the market portfolio;
\item $\vec \beta_m$ is the $N \times 1$ vector of the factor loadings
of the market factor;
\item $\vec \phi$ is the random $N \times 1$ vector of risk factors
$\phi_i$ which are assumed to have zero mean ($\E\[\phi_i\]=0$), unit variance, are
uncorrelated with each other and with $r_m$;
\item $B$ is the $N \times q$ matrix of factor loadings;
\item $\vec \epsilon$ is the random $N \times 1$ vector of disturbance
terms with zero average $\E \[\vec \epsilon\] = \vec 0$ and covariance
matrix $\Omega = \E \[ \vec \epsilon \cdot \vec \epsilon\]$. 
The disturbance terms are
assumed to be uncorrelated with the market return $r_m$ and the factors
$\phi_i$.
\end{itemize}
It would be natural to assume that (i) $\Omega$ is diagonal in order to have 
the $i^{th}$ contribution of $\vec \epsilon$ embodying the specific risk contribution
to the $i^{th}$ asset but, as we shall see in the sequel, the internal consistency
condition makes this impossible and forces the disturbances $\vec \epsilon$ to be
correlated. A weaker hypothesis on $\Omega$ would be that (ii) all its
eigenvalues are uniformly bounded from above by some constant $\lambda$
(i.e., the bound is independent of the size of the economy:
$\displaystyle\forall N, \max_{||x||=1} x' \Omega x \le \lambda$). This implies
that the covariance matrix of the stock returns defined as
\be
\Sigma = \E \[ \(\vec r - \vec \alpha \) \(\vec r - \vec \alpha \)'\]
= \vec \beta \vec \beta' \cdot \Var r_m + BB' + \Omega,
\ee
where the prime denotes the transpose operator, has an approximate $q+1$
factor structure, according to the definition in
\citeasnoun{Chamberlain1983} and \citeasnoun{CR1983}. But these two
assumptions (i) and (ii) are in fact equivalent, as shown by \citeasnoun{GT85}.
Indeed, a simple repackaging of the $N$ security returns into $N$ new
returns constructed by forming $N$ portfolios of the primitive assets
allows one to get a new formulation of expression (\ref{eq:capm}) with mutually
uncorrelated disturbance terms.

To understand why the disturbance terms cannot be uncorrelated, let us
first denote by $\vec w_m$ the vector of the weights of the market
portfolio. Accounting for the fact that the market factor is
itself built upon the universe of assets that it is supposed to explain,
the model must necessarily fulfill the internal consistency relation
\be
r_m = \vec w_m' \cdot \vec r.
\label{mvgmlsl}
\ee
Left-multiplying (\ref{eq:capm}) by $\vec w_m'$, the internal consistency condition
(\ref{mvgmlsl}) implies the following relationship 
\be
\label{eq:market_bis}
 \left[{\vec w'}_m \cdot {\vec
\beta} -1 \right] \cdot \( r_m - \E\[r_m\]\)+  w_m' B \vec \phi + {\vec w'}_m \cdot {\vec \epsilon}=0~.
\ee
Then, by our assumption of absence of correlation between $r_m$, $\vec
\phi$ and $\vec \epsilon$, it follows trivially that\footnote{ Right
multiplying equation (\ref{eq:market_bis}) by $\vec \epsilon'$ and
taking the expectation, given that the return on the market portfolio,
the factors $\vec \phi$ and the disturbance terms $\vec \epsilon$ are
uncorrelated, we obtain that ${\vec w'}_m \cdot \Omega  =0$. Then, right
multiplying ${\vec w'}_m \cdot \Omega  =0$ by ${\vec w}_m$ gives $0
={\vec w'}_m \cdot \Omega \cdot {\vec w}_m = w_m' \cdot
{\rm E}[{\vec \epsilon} \epsilon'] \cdot w_m = {\rm E}[(w_m' \cdot
{\vec \epsilon}) \cdot (w_m' \cdot {\vec \epsilon})'] = {\rm
E}[[w_m' \cdot \vec \epsilon]^2]$, hence the result (\ref{eq:const_eps}).}
\be
\label{eq:const_eps}
\vec w_m' \cdot \vec \epsilon=0 \qquad \text{almost surely},
\ee
while
\be
\vec w_m' \cdot \vec \beta = 1\quad  \text{and} \qquad \vec w_m' B = 0~.
\label{mngjmdkl}
\ee

Several authors have pointed out a consequence of the internal
consistency condition that the market portfolio is made of (or can be
replicated by) the assets they are intended to explain
\cite{Fama1973,Sharpe1990}. An {\it a priori} important consequence of
this internal consistency condition is the breakdown of the standard
assumption of independence (or, at least, of the absence of correlation)
between the non-systematic components of the returns of securities. In
other words, the standard factor
model decompositions assume that
the disturbance terms for security $i$ are uncorrelated with the
comparable components for security $j$. But, this cannot be strictly the case
as can be seen from the above derivation.
This presence of correlations between the disturbance terms 
may {\it a priori} pose problems in the
pricing of portfolio risks: only when the disturbance terms
can be averaged out by diversification can one conclude that the only 
non-diversifiable risk of a portfolio is born out by the contribution of the
market portfolio which is weighted by the beta of the portfolio under
consideration. Previous authors have suggested that this is indeed what
happens in economies in the limit
of a large market $N \to \infty$, for which the
correlations between the disturbance terms vanish asymptotically and the internal consistency
condition seems irrelevant.
For example, while \citeasnoun[footnote 13]{Sharpe1990} concluded
that, as a consequence of equation (\ref{eq:const_eps}), at least two of the
disturbances, say $\epsilon_i$ and $\epsilon_j$, must be negatively correlated, he
suggested that this problem may disappear in economies with infinitely many securities. 
Actually, we show below that this apparently quite reasonable line of
reasoning does not tell the whole
story: even for economies with infinitely many securities, when the companies exhibit
a large distribution of sizes as they do in reality, the constraint (\ref{eq:const_eps})
can lead to the important consequence that the risk born out by an investor holding a 
well-diversified portfolio does not reduce to the market risk in the limit of a 
very large portfolio, as usually believed. A significant proportion 
of ``specific risk'' may remain which cannot be diversified away by a
simple aggregation of a very large number of assets.

\subsection{Correlation structure of the disturbance terms}

The fact that the disturbance terms $\vec \epsilon$ in the market model
(\ref{eq:capm}) are correlated according to the condition (\ref{eq:const_eps})
means that there exists at least one common ``factor'' $f$ to the $\epsilon$'s, so
that $\vec \epsilon$ can be expressed as
\be
\vec \epsilon = \vec \gamma \cdot f + \vec \eta~,
\label{eq:smeori}
\ee
where $\vec \gamma$ is the vector of loading of the factor
$f$\footnote{With this representation, we avoid the case where the
explaining factor -- here the market portfolio -- could be replicated by
a single traded asset. Indeed, in such a case, the replicating portfolio
would be concentrated on one single asset, say the first one, so that
the internal consistency condition would read $\epsilon_1 =0$ without
any other constraint on the $\epsilon_i$, $i>1$.}. The factor $f$ could
be chosen a priori such as to explain one of the many anomalies reported in the
previous section. But, as recalled, we want to move away from this logic
of invoking macro-economic, firm-specific or behavioral factors. We prefer 
to focus on the parsimonious single market factor model, and just account for the lack
of diversification of the market portfolio which calls for a
diversification premium. As a bonus, we will see that this strategy
turns out to provide a fundamental basis of explaining a significant
part of the pricing anomalies.
Our only requirement is that the covariance matrix of
$\vec \epsilon$ exhibits an eigenvalue that goes to infinity in the
limit of an infinite economy, when $H_N$ does not go to zero.
In contrast, when $H_N$ goes to zero as $N \to \infty$, the 
largest eigenvalue should remain bounded. This requirement derives
simply from the results of \citeasnoun{Chamberlain1983} and
\citeasnoun{CR1983}, who have linked the existence of $K$ unbounded
eigenvalues (in the limit $N \to \infty$) of the covariance matrix of
the asset returns to a unique approximate factor structure, such that
the $K$ associated eigenvectors converge and play the role of $K$ factor
loadings.

For simplicity, we choose $\vec \eta$ to be a vector of {\em
uncorrelated} residuals with zero mean\footnote{It should be enough to
assume that all the eigenvalues of the covariance matrix of $\vec \eta$
are positive and uniformly bounded by some positive constant
\cite{GT83}.}. Since $\vec w_m'
\vec \epsilon =0$, $f$ and $\vec \eta$ are not independent from one
another. More precisely, we have
\be
\label{eq:dtpio}
f=- \frac{\vec w_m' \vec \eta}{\vec w_m' \vec \gamma},
\ee
provided that $\vec w_m' \vec \gamma \neq 0$; if not, the random vector
$\vec \eta$ would have to satisfy $\vec w_m' \vec \eta =0$, which
contradicts our assumption of an absence of correlations between the
components of $\vec \eta$. Therefore, in this
framework, $f$ is not actually a factor -- it should be uncorrelated
with $\vec \eta$ if it was -- but is rather an ``endogenous'' factor. The market model
(\ref{eq:capm}) then becomes
\be
\label{eq:capm2}
\vec r = \vec \alpha + \vec \beta \cdot \[ r_m - \E\[r_m\]\] + \vec \gamma \cdot f + \vec \eta,
\ee
with
\begin{itemize}
\item $\Cov(r_m,f)=\Cov(r_m,\vec \eta)=0$, as the result of the absence of correlation between $r_m$ and $\vec \epsilon$,
\item $\Var\vec \eta=\Delta$, where $\Delta$ is a diagonal matrix,
\item $\Var f = \frac{\vec w_m' \Delta \vec w_m}{\(\vec w_m' \vec \gamma\)^2}$, and
\item $\Cov(f,\vec \eta)=-\frac{1}{\vec w_m' \vec \gamma} \cdot \vec w_m' \Delta$.
\end{itemize}

In order to understand and illustrate the relevance and the limits of the
assertion according to which the existence of correlations between two
disturbance terms $\epsilon_i$ and $\epsilon_j$ should be negligible in
an infinite size market \cite{Fama1973,Sharpe1990}, let us now evaluate
their typical magnitude. To simplify the notations, let us rescale
without loss of generality the vector $\vec \gamma$ by $\vec w_m' \vec
\gamma $, so that
the relation (\ref{eq:dtpio}) becomes
\be
f=- \vec w_m' \vec \eta,
\ee
with $\vec w_m' \vec \gamma =1$. 
The covariance matrix $\Omega$ of $\vec \epsilon$ is 
\be
\label{eq:fgjk}
\Omega = \(\vec w_m' \Delta \vec w_m\) \vec \gamma \vec \gamma' - \vec
\gamma \vec w_m' \Delta - \Delta \vec w_m \vec \gamma' + \Delta,
\ee
and the correlation between $\epsilon_i$ and $\epsilon_j$ ($i\neq j$) is
\be
\rho_{ij}=\frac{\(\vec w_m' \Delta \vec w_m\) \gamma_i \gamma_j -
\gamma_i w_{m,j} \Delta_{jj} - \gamma_j w_{m,i}
\Delta_{ii}}{\sqrt{\[\(\vec w_m' \Delta \vec w_m\) \gamma_i^2 -
2\gamma_i w_{m,i} \Delta_{ii} + \Delta_{ii}\] \cdot
\[\(\vec w_m' \Delta \vec w_m\) \gamma_j^2 - 2\gamma_j w_{m,j} \Delta_{jj} + \Delta_{jj}\]}}~.
\label{mgms}
\ee
For illustration purpose, let us assume that all the $\gamma_i$'s are equal
to one (the condition $\vec w_m' \vec \gamma=1$ is then automatically satisfied
from the normalization of the weights $\vec w_m$) and that
$\Delta_{ii} = \Delta$ for all $i$'s. The cumbersome relation (\ref{mgms})
simplifies into
\bea
\rho_{ij} &=& \displaystyle \frac{H_N - w_{m,i} - w_{m,j}}{\sqrt{\( 1+ H_N -
2w_{m,i}\)\( 1+ H_N - 2w_{m,j}\)}},\\
&=& \displaystyle \frac{H_N}{1+H_N} \cdot \(1 + O(w_{m,i(j)}/H_N)\)~.
\label{mkigir}
\eea
Then, expression (\ref{mkigir}) shows that,
provided that the market portfolio is sufficiently well-diversified,
namely provided that the weight of each asset and the concentration index goes to zero 
in the limit of a large market ($N \to \infty$), the
correlations $\rho_{ij}$ between any two disturbance terms goes to zero as
usually assumed. However, the largest eigenvalue of the correlation
matrix, associated with the (asymptotic) eigenvector $\vec 1 = \(1,1,
\dots,1\)'$, is $\lambda_{\max,N}\simeq N \cdot \frac{H_N}{1+H_N}$ and
goes to infinity, as the size of the economy growths unbounded, as soon
as $H_N$ goes to zero more slowly than $1/N$. This clearly shows that
the correlations between the disturbance terms are not necessarily
negligible.

The question, that we now have to address, is whether these weak
correlations may challenge the usual assumption that well-diversified
portfolios do not bear additional non-diversified sources of risks. 
For this, let us consider a well diversified portfolio $ \vec
w_p$, {\it i.e.}, a portfolio such that $||w_p||^2 \rightarrow 0$ as the size
of the economy goes to infinity. From equation (\ref{eq:fgjk}), 
the residual variance of this portfolio, namely the part of the variance of the portfolio that cannot be ascribed to systematic risk factors, reads
\be
w_p'\Omega w_p = \(\vec w_m' \Delta \vec w_m\) \(\vec \gamma \vec
w_p'\)^2 - 2 \(\vec w_m' \Delta \vec w_p\) \(\vec \gamma' \vec w_p\) + 
\vec w_p'\Delta \vec w_p' ~.
\ee
In addition to our previous hypothesis that $\Delta$ is a diagonal matrix, we assume that its entries are uniformly bounded from below by some positive constant $c_1$ and from above by some constant $c_2 <\infty$ and that $|\vec \gamma \vec w_p'|$ is uniformly bounded from below by some
positive constant $c'$ and from above by some finite constant $c''$ (this is
the case, for instance, when one considers $\vec \gamma = \vec 1$, which
is compatible with the requirement $w_m' \cdot \vec \gamma =1$ assumed
in the representation (\ref{eq:fgjk})). Then
\bea
\vec w_p'\Delta \vec w_p' &\le& c_2 \cdot ||\vec w_p||^2 \rightarrow 0,\\
\left|\(\vec w_m' \Delta \vec w_p\) \(\vec \gamma' \vec w_p\)\right|
&\le& c_2 \cdot c'' \cdot ||w_m|| \cdot ||w_p|| \rightarrow 0,
\eea
and
\be
c_1 \cdot c' \cdot ||w_m||^2 \le \(\vec w_m' \Delta \vec w_m\) \(\vec \gamma \vec w_p'\)^2 \le c_2 \cdot c'' \cdot ||w_m||^2,
\ee
so that
\be
\label{eq:kfdyiy}
w_p'\Omega w_p \sim K \cdot H_N, \quad K>0, ~~\text{as}~~~ N\rightarrow \infty.
\ee
Therefore, the residual variance $\vec w_p'\Omega \vec w_p$ of any
``well-diversified portfolio'' $\vec w_p$  goes to zero, as the size $N$
of the economy goes to infinity, {\em if and only if} the concentration
index $H_N$ of the market portfolio goes to zero. In the case of a real
economy, section~\ref{sec1} has shown that the Herfindahl index $H_N$ 
of the market portfolio goes to zero but at the particularly slow decay
rate of $1/(\ln N)^2$. As a consequence, the residual
variance may still account for a significant part of the total
portfolio variance. We will give a numerical example in the next
paragraph providing a more precise statement concerning the behavior of the residual
variance of the equally-weighted portfolio.

\subsection{Asymptotic behavior of the variance of the excess return of
the equally-weighted portfolio}

In order to investigate more precisely the impact of the correlations between
the disturbance terms induced by the condition of internal consistency
on the variance of the  returns of a ``well-diversified'' portfolio, we
consider first the simple case of the equally-weighted portfolio whose
composition is given by the vector
$\vec w_e = \frac{1}{N} \vec 1$. Algebraic manipulations yield
\be
\label{eq:gpok}
\Var r_e = {\beta_e}^2 \cdot \Var r_m + {\bar \gamma_N}^2 \cdot
\frac{\sum_{i=1}^N S_i^2 \Delta_{ii}}{\(\sum_{i=1}^N S_i \gamma_i\)^2} -
2 \bar \gamma_N \frac{1}{N} \cdot \frac{\sum_{i=1}^N S_i
\Delta_{ii}}{\sum_{i=1}^N S_i \gamma_i}+ \frac{1}{N}\( \frac{1}{N}
\sum_{i=1}^N \Delta_{ii}\),
\ee
where $r_e$ denotes the return on the equally-weighted portfolio and
$\beta_e$ its beta with the market factor. We have reintroduced the
explicit dependence on the term $\sum_{i=1}^N w_{m,i} \gamma_i$ (no more
assumed to be scaled to the value $1$) and have explicited the fact
that the market weight of firm $i$ is $w_{m,i}=S_i/\sum_{i=1}^n S_i$.

Two of the four terms in the right-hand-side (r.h.s.) of expression
(\ref{eq:gpok}) are standard.
The first term ${\beta_e}^2 \cdot \Var r_m$ is the traditional contribution
of the market risk factor weighted by the beta of the portfolio.
The last rightmost term in the r.h.s. of (\ref{eq:gpok}) represents the usual
contribution of the diversifiable risk of the portfolio when one assumes that the
disturbance terms are uncorrelated and therefore represents the
specific sources of risk. The two other terms are new and result
from the existence of correlations between the disturbances. In the absence
of such correlations, ${\bar \gamma}$ would be zero and these two terms disappear.

Assuming that the $\Delta_{ii}$'s are $N$ iid (positive) random
variables with finite expected value $\E\[\Delta_{ii}\]  <
\infty$, we get that $\frac{1}{N^2} \sum_{i=1}^N \Delta_{ii}$ and
$\frac{1}{N} \(\bar \gamma_N \cdot \frac{\sum_{i=1}^N S_i
\Delta_{ii}}{\sum_{i=1}^N S_i \gamma_i}\)$ are $O_p\(1/N\)$,
irrespective of the fact that the distribution of firm sizes admits or
does not admit a finite mean\footnote{The term within the parentheses
converges in law either to zero, if $\E[S]<\infty$, or to some non
degenerated distribution,
if $S$ is regularly varying with tail index less than one.}. 
This implies that, in the limit of large $N$, the third and fourth
terms in the r.h.s. of expression (\ref{eq:gpok}) can be 
neglected, leading to 
\be
\Var r_e = {\beta_e}^2 \cdot \Var r_m + {\bar \gamma_N}^2 \cdot
\frac{\sum_{i=1}^N S_i^2 \Delta_{ii}}{\(\sum_{i=1}^N S_i \gamma_i\)^2} +
O_p(1/N).
\label{ngjmlcm,q}
\ee
The fact that the fourth term in expression (\ref{eq:gpok}) disappears
in the limit $N \to \infty$ is not surprising since it recovers the standard
result on the diversification of the idiosyncratic risks. More interestingly,
the fact that the third term in 
(\ref{eq:gpok}) also goes to zero as $1/N$ means that
it does not introduce (in the limit of a large market) an additional risk worth 
considering.

Proposition~\ref{prop2} below reveals
through expression (\ref{ngjmlcm,q}) that the only significant additional
contribution to the risks of the equally-weighted portfolio stems from the term
\be
\label{eqVarf}
{\bar \gamma_N}^2 \cdot \frac{\sum_{i=1}^N S_i^2
\Delta_{ii}}{\(\sum_{i=1}^N S_i \gamma_i\)^2}~,
\ee
which is nothing but the variance (conditional of the $\gamma_i$'s and
the $S_i$'s) of the term $\bar \gamma_N \cdot f$ resulting from the expression of
the market model (\ref{eq:capm2}).

By the same kind of arguments as in Proposition~\ref{prop3a}, we get that
the contribution (\ref{eqVarf}) exhibits three
different behaviors. Either the variance of the distribution of
firm sizes is finite and the term (\ref{eqVarf}) goes to zero has $1/N$,
or only the mean of the distribution of firm sizes exists and 
the  term (\ref{eqVarf}) goes to
zero at a much slower rate or, finally, if the mean of the distribution
of firm sizes does not exist, the additional risk term (\ref{eqVarf}) 
converges to some finite positive value. More precisely, we can state the following results:
\begin{proposition}
\label{prop2}
Assuming that the $\gamma_{i}$'s are iid random variables such that
$\E\[|\gamma|\] < \infty$, and that the $\Delta_{ii}$'s are iid positive
random variables such that $\E[\Delta_{ii}] = \bar \Delta < \infty$, the asymptotic behavior
of the variance of the equally-weighted portfolio is the following:
\begin{enumerate}
\item provided that $\E[S^2] < \infty$,
$$\Var r_e = {\beta_e}^2 \cdot \Var r_m + O_p(1/N),$$
\item provided that $S$ is regularly varying with tail index $\mu=2$ and
$s^\mu \cdot \Pr\[S>s\] \rightarrow c > 0$, as $s$ goes to infinity,
$$
\Var r_e = {\beta_e}^2 \cdot \Var r_m + 
\frac{c \cdot \bar \Delta}{\E\[S\]^2} \frac{\ln N}{N} + o_p(\ln N /N),
$$
\item provided that $S$ is regularly varying with tail index $\mu \in
(1,2)$ and $s^\mu \cdot \Pr\[S>s\] \rightarrow c > 0$, as $s$ goes to
infinity,
$$
\Var r_e = {\beta_e}^2 \cdot \Var r_m +
\[\frac{\pi c \E\[\Delta^{\mu/2}\]}{2 \Gamma \(\frac{\mu}{2}\) \sin \frac{\mu \pi}{4}}\]^{2/\mu}
\frac{1}{\E\[S\]^2}
\cdot\frac{1}{N^{2-2/\mu}}
\cdot \xi_N + o_p\(\frac{1}{N^{2-2/\mu}}\)~,
$$ 
where $\xi_N$ is a
sequence of positive random variables with stable limit law
$S(\mu/2,1)$,
\item provided that $S$ is regularly varying with tail index $\mu = 1$
and $s^\mu \cdot \Pr\[S>s\] \rightarrow c > 0$, as $s$ goes to infinity,
$$\Var r_e = {\beta_e}^2 \cdot \Var r_m +
\frac{\pi~ \E \[\Delta^{1/2}\]^{2}}{2}
\frac{\E\[\gamma\]^2}{\E\[|\gamma|\]^2}  \frac{1}{\ln^2 N}\cdot \xi_N +
o_p\(1/\ln^2 N\),
$$
where $\xi_N$ is a sequence of positive random variables with stable limit
law $S(1/2,1)$,
\item provided that $S$ is regularly varying with tail index $\mu \in(0,1)$ and $s^\mu \cdot \Pr\[S>s\] \rightarrow c > 0$, as $s$ goes to infinity, 
$$
\Var r_e = {\beta_e}^2 \cdot \Var r_m + \E \[\Delta^{\mu/2}\]^{2/\mu}
\frac{\E\[\gamma\]^2}{\E\[|\gamma|^{\mu}\]^{2/\mu}}
\frac{4}{\pi^{1/\mu}}\[ \Gamma\(\frac{1+\mu}{2}\) \cos \frac{\pi
\mu}{4}\]^{2/\mu}
\cdot \frac{\xi_N}{\zeta_N^2} + o_p(1),$$
where $\xi_N$ and $\zeta_N$ are two
sequences of strongly correlated\footnote{see footnote~\ref{foot}.} positive random variables that converge in law to $S(\mu/2,1)$  and $S(\mu,\beta_\gamma)$ 
with $\beta_\gamma=\frac{\E \[ \gamma^\mu \cdot 1_{\gamma>0}\] - \E \[
 |\gamma|^\mu \cdot 1_{\gamma<0} \]}{ \E \[ |\gamma|^\mu \]}$ respectively.
\end{enumerate}
\end{proposition}

\begin{figure}
\begin{center}
\includegraphics[width=12cm]{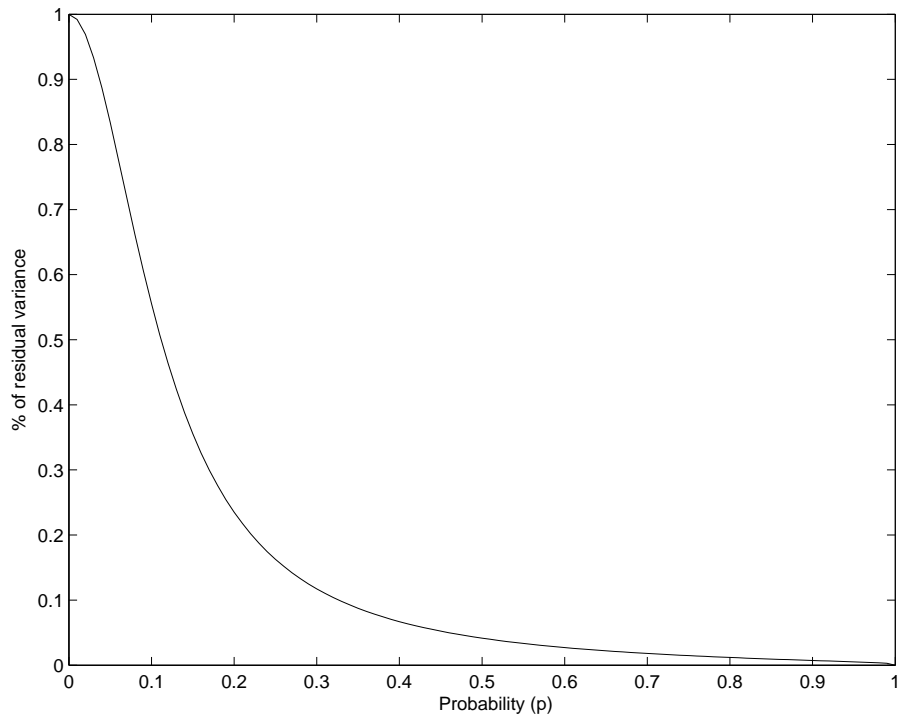}
\end{center}
\captionsetup{style=default,labelfont=bf, labelsep=period}
\caption{\label{figResVar} {\bf Contribution of the residual variance to
the total variance}. The figure shows the probability $p$ to reach or
exceed a given contribution level, in percentage, of the residual
variance to the total variance of the return on the equally weighted
portfolio in a market with 7000-8000 traded assets and with a distribution of
firm sizes given by Zipf's law ($\mu=1$).
}
\end{figure}

Focusing on the case where $\mu$ is equal (or close) to one, as in real 
markets, proposition~\ref{prop2} tells us that the asymptotic behavior
of the variance
of the equally weighted portfolio is given by
\be
\label{mjbjvw}
\Var r_e = {\beta_e}^2 \cdot \Var r_m +
\frac{\pi \cdot \E\[\gamma\]^2 \cdot \E \[\Delta^{1/2}\]^2}{2 \cdot
\E\[|\gamma|\]^2 \cdot \( \ln N\)^2} \cdot \xi_N +
o_p\(1/\( \ln N\)^2\),
\ee
where $\xi_N$ is a sequence of positive random variables with stable
limit law $S(1/2,1)$ whose density is given by (\ref{eq:LevyLaw}).
Expression (\ref{mjbjvw}) implies that the variance of the equally-weighted
portfolio, while asymptotically proportional to the variance of the
market portfolio, receives a significant contribution due to the
internal consistent condition together with the Zipf distribution of company
sizes. This additional contribution decays to zero extremely slowly with the number $N$ of
companies in the economy. For instance, (i) assuming that the variance
$\Delta_{ii}$ of the residuals $\eta_i$ is the same for all of them and,
{\it a priori}, of the same order as the variance of the market return:
$\Delta_{ii} \sim \Var r_m$, (ii) considering that the ratio
$\frac{\E\[\gamma\]^2}{\E\[|\gamma|\]^2}$ is of the order of one and
(iii) accounting for the fact that the median value of the L\'{e}vy law is
approximately equal to $2.198$, the additional term is typically of the
order of $\frac{\pi}{\(\ln N\)^2} \cdot \Var r_m$. So, assuming that
$\beta_e$ is about one and considering a market where $7000$ to $8000$
assets are traded\footnote{These figures are compatible with the number
of stocks currently listed on the Amex, the Nasdaq and the Nyse.},
the typical amplitude of the additional term represents $5\%$ of
the total variance of the equally-weighted portfolio. More precisely, 
in one case out of two, the contribution of the additional
term is larger than $5\%$ of the total variance. 
Figure~\ref{figResVar} presents the probability to reach or exceed a given level
for the contribution of the residual variance to the total
variance, in an economy with 7000-8000 traded assets.
In one case out at four ($p=0.25$), the contribution of the residual
variance to the total variance is larger than $15\%$; in one
case out ten ($p=0.1$), it represents more than $50\%$.

\vspace{0.5cm}
\centerline{\it [ Insert Figure~\ref{figResVar} about here]}
\vspace{0.5cm}

\subsection{Relation with the concentration of the market portfolio}

The variance of the term $\bar \gamma_N \cdot f$ given by
(\ref{eqVarf}) cannot be easily related to observable market variables
since it is a mixture of the firm sizes (which are observable) and of the
not directly accessible underlying variables $\gamma_i$'s and $\Delta_{ii}$'s which
describe the correlation structure of the disturbances $\vec
\epsilon$ in the model (\ref{eq:capm}).
Nevertheless, as a consequence of the
assumption that the $\gamma_i$'s and $\Delta_{ii}$'s have finite
expectations, the behavior of the term $\frac{\sum_{i=1}^N S_i^2
\Delta_{ii}}{\(\sum_{i=1}^N S_i \gamma_i\)^2}$ is the same as that
of $\frac{\sum_{i=1}^N S_i^2}{\(\sum_{i=1}^N S_i\)^2}$ which is
nothing but the Herfindahl index $H_N$ of the market portfolio since
\be
H_N \equiv \sum_{i=1}^N w_{m,i}^2 = \frac{\sum_{i=1}^N S_i^2}{\(\sum_{i=1}^N S_i\)^2}~.
\ee

In fact, propositions~\ref{prop3a} and \ref{prop2} are closely related. Loosely
speaking, these two propositions can be summarized as follows
\be
\label{eq:opf}
\Var r_e \stackrel{law}{\simeq} \beta_e^2 \cdot \Var r_m + K_\mu \cdot H_N,
\ee
where
\be
K_\mu = \left\{
\begin{array}{lr}
\displaystyle \E\[\Delta\], & \mu \ge 2,\\
\displaystyle \E\[\Delta ^{\mu/2}\]^{2/\mu}, & 1< \mu <2,\\
\displaystyle \E\[\Delta ^{\mu/2}\]^{2/\mu} \cdot
\frac{\E\[\gamma\]^2}{\E\[|\gamma|^\mu\]^{2/\mu}}, & \mu \le 1.\\
\end{array}
\right.
\ee

Expression (\ref{eq:opf}) has a simple intuitive meaning based upon the
standard interpretation of the Herfindahl index as the inverse of the
effective number of assets of a portfolio, if this portfolio was
well-diversified (in fact, equally-weighted). Indeed, considering an
equally-weighted portfolio made of $n$ assets, its Herfindahl index is
$H=1/n$. Conversely, given a portfolio whose Herfindahl index is $H$,
its effective number of assets, defined as the number of assets of an
equally-weighted portfolio with the same value $H$ of the Herfindahl
index, is $n_{eff}=1/H$. Therefore, considering that the real market is
not made of $N$ $(\simeq 7000-8000)$ assets but actually of $N_{eff}=1/H_N$
$(\simeq 20-25)$ effective assets, equation (\ref{eq:opf}) expresses the
variance of the equally-weighted portfolio as the sum of two terms: the
first one gives the variance of the portfolio resulting from the
exposition to the market risk $\beta_e^2 \cdot \Var r_m$ while the
second one represents the residual variance of the $N_{eff}=1/H_N$
assets. The constant $K_\mu$ appears as the average residual variance of
the $N_{eff}$ assets. Thus, when the market portfolio is
well-diversified, $H_N$ goes to zero, or equivalently, the number of
effective assets goes to infinity so that, by virtue of the law of large
numbers, the residual variance $K_\mu/N_{eff}$ goes to zero. In contrast,
when the market portfolio is concentrated on a few assets,
$H_N$ does not go to zero, the number of effective assets remains finite
in the limit of an infinite economy and the residual variance does not
go to zero.

For illustration purpose, we discuss in turn three cases.
First, both propositions~\ref{prop3a} and
\ref{prop2} show that the concentration
index $H_N$ and the variance of $f$ are of the order of $1/N$, like the
last two terms in the r.h.s. of expression (\ref{eq:gpok}), provided that the variance of the
distribution of firm sizes is finite. 
As a consequence, for such
distributions of firm sizes, the market portfolio is well diversified
insofar as the concentration index is of the same order as the inverse
of the number of assets in the portfolio. As a consequence, there is no additional
non-diversifiable risk and, in the limit of a large market, we have
\be
\Var r_e = \beta_e^2 \cdot \Var r_m + O_p(N^{-1}).
\ee
Let us consider the example of a distribution of firm
sizes given by a Gamma law $\Gamma(r,\lambda)$. In such a case, it is
well-known that the joint distribution of $\{w_{m,i}\}_{i=1}^{N-1}$ is
a multivariate Beta law with parameter $r$ \cite{Mosimann1962}, which yields 
\be
\E[H_N] = \frac{r+1}{r \cdot N+1},
\ee
in accordance with the fact that $H_n = \frac{1}{N} +o_p(1/N)$.

Second, if the distribution of firm sizes admits only a finite mean value
and, in addition, is regularly varying at infinity with a tail index $\mu
\in (1,2)$, the propositions~\ref{prop3a} and
\ref{prop2} state that both
the concentration index and the variance of $f$ are of the order of
$1/N^{2\(1-1/\mu\)}$. As a consequence, 
the contribution to the total risk due to the second term 
in the r.h.s. of (\ref{eq:gpok}) decays to zero much slower than the decay
$\sim 1/N$ of the two last terms. Then
\be
\Var r_e =  \beta_e^2 \cdot \Var r_m +
\frac{C}{N^{2\(1-\frac{1}{\mu}\)}} + O_p(N^{-1}), \qquad
\text{for some } C>0.
\ee
As an example, if the tail index of the distribution of firm sizes is
$\mu=3/2$, the ratio of the second term in the r.h.s. of (\ref{eq:gpok}) over
the last two terms is of the order of $N^{1/3}$. Therefore, 
assuming that the prefactors of these contributions have 
the same magnitude, the second term is typically $10, 21$ and $46$ times larger than the
last two terms, if one thousand, ten thousands and one hundred
thousand companies are traded on the market.

Finally, if the distribution of firm sizes does not even admit a finite
mean value but is still regularly varying at infinity with a tail index
$\mu \in (0,1)$, propositions~\ref{prop3a} and \ref{prop2} show
that the Herfindahl index and the variance of $f$ converge to
non-degenerated random variables which are the ratio of two positive and
dependent stable random variables:
\bea
H_N &=& H + o_p(1), \quad {\rm with}~ H = \lim_{N\rightarrow \infty}
\frac{S_1^2 + \cdots + S_N^2}{\(S_1 + \cdots + S_N \)^2}, \label{mmlclw}\\ 
\Var f &=&
\sigma_f^2 + o_p(1), \quad {\rm with}~ \sigma_f^2 = \lim_{N\rightarrow
\infty} \frac{\Delta_{11} \cdot S_1^2 + \cdots + \Delta_{NN} \cdot
S_N^2}{\(\gamma_1 \cdot S_1 + \cdots + \gamma_N \cdot S_N \)^2}~,
\eea
so that
\be
\Var r_e = \underbrace{\beta_e^2 \cdot \Var r_m}_{\rm specific~
market~ risk} + \underbrace{\E\[\gamma\]^2 \cdot \sigma_f^2}_{\rm
non-diversified~ risk} + o_p(1).
\label{mghkbwp}
\ee 
The first term in the r.h.s. of (\ref{mghkbwp}) is the non-diversifiable market risk which is
remunerated by the market according to the CAPM formula. The second term
clearly exemplifies the fact that due to (i) the dependence between the
$\vec \epsilon$ resulting from the internal consistency
condition and (ii) the Pareto form of the distribution of the size of
companies, full diversification cannot occur even in the
limit of a market with an infinite number of assets.
Consider the example where the distribution of firm sizes
is the L\'evy law defined by equation (\ref{eq:LevyLaw}).
Using its properties of stability under convolution, the
distribution of the market weights $w_{m,i}$ can be easily obtained. For instance, the density
of the marginal law of $w_i$ is given by
\be
g_N(w)=\frac{N-1}{\pi} \cdot \frac{w^{-1/2}(1-w)^{1/2}}{1 + \[(N-1)^2 -1\]w}~,
\ee
so that $\E \[H_N\] = \frac{1}{2} \cdot \frac{N+1}{N}$, in agreement
with the fifth statement of proposition \ref{prop3a} and (\ref{mmlclw}).

\subsection{Generalization to arbitrary well-diversified portfolios \label{sec:GWD}}

The detailed results obtained until now in
section~\ref{sec:Diversification} refer to one particular portfolio, the
equally-weighted portfolio. This portfolio is interesting because it is
often taken as a reference and as a starting point to more elaborate
allocations by analysts and practitioners. However, from the previous sections,
it seems natural to conjecture that the results summarized in
proposition \ref{prop2} also hold (with suitable adaptation) for the
entire class of well-diversified portfolios as suggested by equation
(\ref{eq:kfdyiy}). By {\em well-diversified} portfolio is meant a
portfolio of $N$ assets whose concentration index goes to zero in the limit of
large $N$. In the particular case where we consider a portfolio $p$,
with weight on asset $i$ given by $w_{p,i}=\alpha_i/N$, where the
$\alpha_i$'s have to sum up to $N$ in order to ensure that the sum of
the fractions of wealth invested in each asset is equal to one and such
that $\frac{1}{N} \sum_{i=1}^N \alpha_i^2$ is uniformly bounded from
above by some finite constant,  the Herfindahl index of $p$ behaves as
\be
H_{p,N} \sim \frac{C}{N}, \qquad \text{as}~ N \rightarrow \infty,
\ee
where $C$ is a positive and finite constant. Then, the variance  of portfolio $p$ reads
\be
\label{eq:hgdlf}
\Var r_p = {\beta_p}^2 \cdot \Var r_m + \E\[\gamma\]^2
\cdot \frac{\sum_{i=1}^N S_i^2 \Delta_{ii}}{\(\sum_{i=1}^N S_i
\gamma_i\)^2}+   o_p(1)~,
\ee
by virtue of the law of large numbers.

This expression shows that the term
$\frac{\sum_{i=1}^N S_i^2 \Delta_{ii}}{\(\sum_{i=1}^N S_i
\gamma_i\)^2}$ (or equivalently the concentration index $H_N$ of the
market portfolio) still controls the decay (or the absence of decay) of
the contribution to the variance in addition to the variance associated with the
correlation of the portfolio $p$ with the market portfolio. Therefore,
we conclude that proposition~\ref{prop2} holds for the entire class of
portfolios whose Herfindahl index decays to zero as $C/N$, for large
$N$. In fact, the result holds for this class of {\em long} portfolios,
{\it i.e.} such that the weights $\alpha_i/N$ sum up to one. In the case
of an {\em arbitrage} portfolio, namely a portfolio whose weights
$\alpha_i/N$ sum up to zero, no additional term appears in the variance
(\ref{eq:hgdlf}).

Finally, when the concentration index of the portfolio under
consideration goes to zero, but at rate slower from $1/N$, 
obtaining a detailed result for the
variance of the portfolio's return involves more complex formulas.
For the present work, equation
(\ref{eq:kfdyiy}) is sufficient to state that, in general,
well-diversified portfolios, of which the equally-weighted portfolio is
just an example, have generally a non-diversified risk which does not
vanish in the limit of large economies, if the distribution of firm
capitalizations is sufficiently heavy-tailed. Therefore, holding a
portfolio with asymptotically vanishing Herfindahl index does not
necessarily diversify away the non-systematic risk.

\section{Discussion}

\subsection{Analysis of synthetic markets generated numerically}
\label{sec:NumSim}

In order to assess the impact of the internal consistency factor 
in real stock markets of finite size, we
present in table~\ref{tab:NumSim} the results of numerical simulations
of synthetic markets with respectively $N=1000$ and $N=10000$
traded assets. We construct the synthetic markets according to model
(\ref{eq:capm2}) so that the only {\em explicit} explaining factor is
the market factor and the
size distribution of the capitalization of firms is the Pareto distribution
\be
\Pr \[S \ge s \] = \frac{1}{s^\mu} \cdot 1_{s\ge1}~.
\ee
We investigate various synthetic markets characterized by different tail
index $\mu$, from $\mu=1/2$ (deep in the heavy-tailed regime), $\mu=1$
(borderline case often referred to as the Zipf law when expressed with
sizes plotted as a function of ranks) to $\mu=2$ (for which the central
limit theorem holds and standard results are expected). It is important
to stress that the results presented in table~\ref{tab:NumSim} are
insensitive to the shape of the bulk of the distribution of firm sizes,
and only the tail $\Pr \[S \ge s \] \sim {s^{-\mu}}$, for large $s$,
matters.

The three values of the tail index $\mu$ equal to $2$, $1$ and
$1/2$ correspond to the three major behaviors of the residual
variance of a ``well-diversified'' portfolio, namely the part of the
total variance related to the disturbance term $\epsilon$ only, given by
proposition~\ref{prop2}:
\begin{itemize}
\item for $\mu = 2$, the residual variance goes to zero as $1/N$, so that the market return should be
the only relevant explaining factor if the the number of traded assets
is large enough;
\item for $\mu = 1$, the residual variance goes very slowly to zero, so
that one can expect a significant contribution to the total risk and a
strong impact of the internal consistency factor $f$ for large (but
finite) market sizes;
\item for $\mu = 1/2$, the residual variance does not go to zero and one
can expect that the contribution of the residual variance to the total
risk remains a finite contribution as the size of the market
increases without bounds.
\end{itemize}

For each value $\mu=2$, $\mu=1$ and $\mu=1/2$, we generate
100 synthetic markets of each size $N=1000$ and $N=10000$ (hence a total
of $3 \times 2 \times 100$ synthetic markets). For each market, 
we construct 20 equally weighted portfolios (randomly
drawn from each market) and we regress their returns on the returns
of the market portfolio
($r_m$), on the returns of the market portfolio and of the internal consistency factor
($r_m, f$), on the returns of the market portfolio and of the (overall) equally weighted
portfolio ($r_m, r_e$), on the returns of the market portfolio and of an arbitrary under-diversified
portfolio ($r_m, r_u$) and on the returns of the market portfolio and of an arbitrary
well-diversified arbitrage portfolio ($r_m, r_a$).
Using the 100 market simulations for each case ($\mu$, $N$),
Table~\ref{tab:NumSim} summarizes the
mean, minimum and maximum values of the coefficient of determination $R^2$ of these five regressions of the 20 equally weighted portfolios.

\vspace{0.5cm}
\centerline{\it [Insert Table~\ref{tab:NumSim} about here]}
\vspace{0.5cm}

For $\mu=2$, as was expected, the market return is the only relevant
factor: it accounts on average for about 95\% and 99\% (for $N=1000$ and
$N=10000$ assets, respectively) of the total variance of the 20
equally-weighted portfolios under considerations. The fact, that the
explained variance increases from 95\% to 99\% when going from $N=1000$
to $N=10000$ assets, results from the standard diversification effect:
for $N=1000$, each of the 20 equally-weighted portfolios are made of
only 1000/20=50 assets compared with 10000/20=500 assets for $N=10000$.
As a confirmation, the minimum and maximum values of the $R^2$
remains very close to their respective mean values.

For $\mu=1$, the market factor explains a much smaller part of the total
variance compared with the previous case (80\% and 88\%, respectively
for $N=1000$ and $N=10000$ assets). As expected, this effect is stronger
for the markets with the smallest number $N=1000$ of traded assets. In
addition, the minimum $R^2$ (1\% and 20\%, resp.)
departs strongly from its mean value. Besides, the
regression on the market factor and the internal consistency factor $f$
(which is readily accessible in the case of a numerical simulation) provides a
level of explanation (95\% and 99\%, respectively) comparable to that of the
case $\mu=2$ for which full diversification of the residual risk occurs. Moreover, 
the equally-weighted portfolio provides the same level of explanation
as $f$ itself. This is particularly interesting insofar as $f$ is not
observable in a real market while the return on the equally-weighted
portfolio can always be calculated, or at least proxied. We find more
generally that any well-diversified portfolio provides overall the same
explaining power. This result is
simply related to the fact that the internal consistency factor $f$ is
responsible for the lack of diversification of `` well-diversified''
portfolios (when $\mu \lesssim 1$) so that the return on any ``well-diversified'' portfolio
$p$ reads $r_p \simeq \alpha_p + \beta_p \cdot r_m + \E\[\gamma\] \cdot f$.
This suggests that the equally-weighted portfolio or any well-diversified
portfolio , in so far as it is strongly sensitive to the internal consistency factor $f$, may act as a good proxy for this factor.

In contrast, the regression on any under-diversified portfolio, while
improving on the  regression performed just using the market portfolio, remains
of lower quality: the gain in $R^2$ is only 5-6\% on average with
respect to the regression on the sole market portfolio, while the gain
in $R^2$ lies in the range 10-15\% when using the equally-weighted
portfolio. Finally, table~\ref{tab:NumSim} shows that the introduction
of an arbitrage portfolio does not improve the regression. This is due
to the fact that arbitrage portfolios are not asymptotically sensitive to
the internal consistency factor $f$ in the large $N$ limit, as recalled in section II-\ref{sec:GWD}.

The same conclusions hold qualitatively for synthetic markets generated
with $\mu=1/2$, with the important quantitative change that the explanatory
power of the market factor does not increase with the market size $N$.
This expresses the predicted property that the internal consistency factor $f$
should have an asymptotically finite contribution to the residual variance 
as the size of the market increases without bounds.

Finally, our numerical tests confirm that the distributional properties of the $\gamma$'s
(the factors loading of the residuals on the internal consistency factor $f$)
have no significant impact on the results of the simulation, provided that
$\E\[|\gamma|\] < \infty$.

\subsection{Consequences for the Arbitrage Pricing Theory and the
standard pricing anomalies}

In his article establishing the arbitrage pricing theory,
\citeasnoun[p. 347]{Ross1976} explicitly assumes that the disturbance
terms in the factor model (\ref{eq:capm}) are ``mutually
stochastically uncorrelated,'' which is inconsistent with the constraint
(\ref{eq:const_eps}) if we assume that the factors (or at least some of them)
can be replicated by assets portfolios. Indeed, the derivation of the
APT results from the construction of a {\em well-diversified} arbitrage
portfolio (step 1 in \citeasnoun[p. 342]{Ross1976}) chosen so as to have
no systematic risk (step 2). The fact that this arbitrage portfolio is
well-diversified is important because it is at the basis of the argument
for the diversification of the specific risk of the arbitrage
portfolio in the limit of a large number of assets (law of large
numbers), which conditions the results of steps 3 and 4 in \citeasnoun{Ross1976}.
Unfortunately, as shown in section~II-E, if
one of the factors can be replicated by a portfolio whose weights are
distributed according to a sufficiently fat-tailed distribution, the
specific risk of this portfolio cannot be diversified away even if it is
a well-diversified portfolio, as defined in section~II-E. 
In that case, the conclusion resulting from steps 3 and 4 in
\citeasnoun{Ross1976} breaks down.

Alternatively, we can say that the residual risks exhibit too strong
correlations. This problem has been tackled by many authors. In particular, 
\citeasnoun{Chamberlain1983} and \citeasnoun{CR1983} have developed the
appropriate formalism to deal with it, while \citeasnoun{Stambaugh1982} and
\citeasnoun{Ingersoll1984}
have provided sharp pricing bounds in the presence of correlation between the
error terms. Basically, when all the
eigenvalues of their covariance matrix remains bounded as more
and more assets are added to the market until its size goes to infinity,
the ATP holds. In contrast, when several eigenvalues grow without
bound, the factors associated with these eigenvalues must be split off
from the residuals and considered as new explaining factors that should be
priced. This argument is at the basis of the choice of the specification
(\ref{eq:smeori}) of the dependence structure of the disturbances of our
market model. Therefore, if we explicitly include our additional
internal consistency risk factor $f$ in the analysis, the original
derivation of Ross' results still holds, as shown by
\citeasnoun{Chamberlain1983}. Indeed, a key
technical assumption for the APT to hold is that the $\epsilon_i$'s (in
equation (\ref{eq:capm})) are ``sufficiently
independent to ensure that the law of large numbers holds'' \cite[p. 342]{Ross1976} 
and, as explained in the previous sections, this condition
breaks down. Nonetheless, this condition holds for the residuals
$\eta_i$ defined by equations (\ref{eq:smeori}-\ref{eq:capm2}). Then,
for the one factor model (\ref{eq:capm2}), the following result holds:
\begin{proposition}
\label{prop4}
Consider a market where $N$ assets are traded and for which the internal
consistency condition (\ref{eq:const_eps}) holds,
so that the returns of the set of assets obey the following dynamics:
$\vec r = \E \[ \vec r\] + \vec \beta \cdot \[r_m - \E \[r_m\]\] + \vec
\gamma \cdot f + \vec \eta$, where $f$ is the (zero-mean) additional
factor resulting from the internal consistency condition and $r_m$ is
uncorrelated with $f$ and the centered disturbance vector $\vec \eta$.
Then, under the usual assumptions required for the APT to hold, 
the expected return on asset $i$ satisfies
\be
\E \[r_i - r_0\] = \beta_i \cdot \E\[r_m - r_0\] + \(\gamma_i - \gamma_m
\cdot \beta_i \) \cdot \E\[r_{icc} - r_0\]~,
\label{kggbnmwbfvm}
\ee
where $r_0$ denotes the risk free interest rate and $\E\[r_{icc}\] \ge
r_0$ is the expected return on any portfolio $\vec w_{icc}$ such that
$\vec w_{icc}' \cdot  \vec \beta =0$, with unit exposure to the factor
$f$ -- {\it i.e.} such that $\vec
w_{icc}' \cdot  \vec \gamma =1$ --  and which is well-diversified in the
sense that the variance $\Var \( \vec  w_{icc} \cdot \vec \eta\)$
goes to zero as the number $N$ of assets goes to infinity. $\gamma_m =
\vec w_m' \cdot \vec \gamma$ is the gamma of the market portfolio.
The index $ _{icc}$ refers to the ``internal consistency condition.''
\end{proposition}

The proof of this result proceeds as follows. Starting
from the model (\ref{eq:capm2}) and following step by step the demonstration
of theorems I and II in \citeasnoun{Ross1976}, we get the asymptotic
result
\be
\E\[\vec r \] = \rho \vec 1 + \lambda_1 \vec \beta + \lambda_2 \vec \gamma,
\ee
where $\rho$, $\lambda_1$ and $\lambda_2$ are three non-negative
constants. Their values are determined by expressing the expected return on
the market portfolio $w_m$, on the portfolio $w_{icc}$ and on any
well-diversified portfolio without any systematic risk. This leads
to identifying $\rho$ with $r_0$, $\lambda_2$ with ${\rm E}\[r_{icc} - r_0\]$ and $\lambda_1$ with
$- \gamma_m\cdot \E \[r_{icc} - r_0\] - r_0$. The quantity $\gamma_m=\vec w_m' \cdot \vec \gamma$ never vanishes, due
to the internal consistency constraint of the model.

Two comments are in order. Firstly, expression (\ref{kggbnmwbfvm}) looks like a standard APT decomposition of the risk premia
of the expected return of a given asset $i$ weighted
by their factor loading, except for one important feature: the risk premium due to the
internal consistency factor has its amplitude controlled by the factor
loading $\gamma_i$ (as usual) {\em corrected} by the unusual  term
$-\gamma_m \beta_i$. In a standard factor decomposition, it is always convenient
to impose $\gamma_m \equiv {\vec w_m}'.{\vec \gamma}=0$ so that
the contribution to the total risk premium due to
any factor is proportional to its corresponding factor loading $\gamma_i$.
In the factor decomposition including the internal consistency factor, this is
intrinsically impossible, as we have stressed above. In this sense, expression (\ref{kggbnmwbfvm})
is not the result of a standard factor decomposition. It is however the
correct decomposition for a one factor model in the presence of the
internal consistency condition, which may lead to the creation of the new
internal consistency factor. The later should in fact be referred to as an
endogenous factor. This decomposition leading to (\ref{kggbnmwbfvm}) is the correct one
in particular to highlight the crucial consequence of the internal consistency
condition in the contribution of the endogenous factor to the total
risk premium of a given asset. As we shall see, the fact that the factor loading $\beta_i$ of
the market portfolio contributes to the amplitude of the risk premium
due to the endogenous factor provides an interesting interpretation of the
book-to-market effect. 

Secondly, in the case where the market portfolio is
well-diversified, the contribution of the additional risk factor $f$
vanishes asymptotically so that the risk premium associated with
this risk factor goes to zero in the limit of an infinitely large
market.

\subsection{Empirical consequences}

The pricing formula given by proposition~\ref{prop4} offers an
interesting new insight into the valuation of asset prices. However, the
direct assessment of the risk premium associated with the 
internal consistency risk factor
ICC is not possible because we do not have {\it a priori} access to it,
so that the practical implementation of this
theoretical framework seems problematic. Nonetheless, if we recall
that the risk premium associated with the additional term 
$\(\gamma_i - \gamma_m \cdot \beta_i \) \cdot \E\[r_{icc} - r_0\]$ 
is due to the lack of diversification of the so-called ``specific risk,'' and that
well-diversified portfolios such that the equally-weighted portfolio are
particularly sensitive to this risk, it seems natural to consider
the return on this portfolio in order to probe the market price of the
non-diversified risk. Besides, the numerical simulations presented in
section~\ref{sec:NumSim} testify to the relevance of this choice.
However, insofar as the equally-weighted portfolio is (by construction)
strongly correlated with the market portfolio, it can be desirable to consider instead the arbitrage portfolio made of a long position in the equally-weighted portfolio and
of a short position in the market portfolio. This arbitrage portfolio
constitutes our proxy for the ICC risk factor and we denote by $r_{icc}(t)$
the time series of its returns. Therefore, this reasoning applied to 
proposition~\ref{prop4} leads us to estimate the following
regression model
\be
r_{i,t} - r_0 = \alpha_i + \beta_i \cdot \[r_m(t) - r_0\] +
\beta^{ICC}_{i} \cdot r_{icc}(t) + \epsilon_i(t)~.
\label{nhnorgp}
\ee
In order to assess the explaining power of the new factor, we also
include in the regression model the two factors SMB and HML of Fama and French (see \citeasnoun{FamaandFrench1993} for the description of the construction of these two portfolios).
We use the monthly excess returns of twenty-five equally-weighted
portfolios sorted by the quintiles of the distribution of sizes and
book-to-market values and the returns of ten value-weighted and equally-weighted industry
portfolios\footnote{We have used the monthly data available on Professor
French's website:
\url{http://mba.tuck.dartmouth.edu/pages/faculty/ken.french/ftp/
25_Portfolios_5x5.zip} for the 25 portfolios sorted by size and
book-to-market,\\
\url{http://mba.tuck.dartmouth.edu/pages/faculty/ken.french/ftp/
10_Industry_Portfolios.zip} for the ten industry portfolios and\\
\url{http://mba.tuck.dartmouth.edu/pages/faculty/ken.french/ftp/F-
F_Research_Data_Factors.zip} for the market factor, the risk-free
interest rate and the two factors SMB and HML.}. Tables~\ref{tab1a} to~\ref{tab3b}
present our results for the period from Jan. 1927 to Dec. 2005.

\vspace{0.5cm}
\centerline{\it [Insert Tables~\ref{tab1a} and~\ref{tab1b} about here]}
\vspace{0.5cm}

Table~\ref{tab1a} presents the parameter estimates of the multi-linear
time series regression of the excess monthly returns of 25
equally-weighed portfolios (sorted by quintiles of the distribution of
sizes -- Small, 2, 3, 4 and Big -- and
by quintiles of the distribution of Book equity to Market equity ratio
-- Low, 2, 3, 4 and High) regressed on the excess return on the market
portfolio, on the two Fama-French factors SMB and HML and on the proxy ICC
for the additional risk factor due to the internal consistency
constraint given by the difference between the return on the
equally-weighted portfolio and the return on the
market portfolio: 
\be
r_{i,t} - r_0 = \alpha_i + \beta_i \cdot \[r_m(t) -
r_0\] + \beta^{ICC}_{i} \cdot r_{icc}(t) + \beta^{SMB}_{i} \cdot
r_{smb}(t) + \beta^{HML}_{i} \cdot r_{hml}(t) +\epsilon_i(t).
\ee 
The figures decorated by one star (resp. two stars) show the cases which
reject the null hypothesis that the factor under consideration is not
significant in the presence of the others at the 5\% (resp. the 1\%)
level. Clearly, the three factors SMB, HML and ICC are, almost always,
significant at the $1\%$ level, suggesting that it is {\it a priori}
useful to consider these three factors together. The regressions on the
four factors provide a very good explanation of the portfolios excess
returns, as witnessed by the $R^2$'s which are larger than, or of the
order of, $90\%$ for most portfolios, except for three extreme cases:
Small-Low, Small-2 and Big-High.

However, these conclusions must be tempered in view of the results
summarized in table~\ref{tab1b} which gives the $R^2$ of the various regressions of the
multi-linear times series of the monthly excess returns of
these 25 equally-weighed portfolios on the market portfolio (Rm), on the
market portfolio and the factor ICC (ICC), on the market portfolio and
the size factor (SMB), on the market portfolio and the book to market
factor (HML), on the market portfolio and the two Fama\&French factors
(HML + SMB),  on the market portfolio, the factor ICC and the size
factor (ICC + SMB), on the market portfolio, the factor ICC and the book
to market factor (ICC + HML) and, finally on all four factors (market, ICC, 
SMB and HML).
The numbers in boldface represent the maximum value of the $R^2$ within the
group of regression with two factors (columns ICC, SMB and HML) and with
three factors (columns HML + SMB, ICC + SMB and ICC + HML) while the
numbers within parenthesis provide the $95\%$ confidence interval of the
$R^2$ obtained by bootstrap \cite{ET1993}.

Several comments are in order. First, for the {\em two-factor} models --
namely the regression models which include the market factor and one of
the factors ICC, SMB or HML -- the internal consistency factor ICC
provides the best explanation in 11 cases out of 25. Second, for the
groups of portfolios within the first three quintiles of the
distribution of sizes, {\it i.e}, Small, 2 and 3, the factor ICC
provides the largest improvement in 10 cases out of 15. Beside, the
improvement provided by the factor ICC is particularly important for the
group of the five portfolios built on the first quintile of the
distribution of size (group ``Small'') with respect to both the size and
the book-to-market factors. Third, based upon the $95\%$ confidence
intervals (figures within parenthesis) obtained by bootstrap, this
improvement is statistically significant with respect to the regression
on the sole market factor\footnote{Note that, {\it a priori}, the quoted
$R^2$ of the linear models are not directly comparable since they
involve different numbers of parameters. In principle, it is thus
necessary to use the adjusted-$R^2$ instead of the raw $R^2$. However,
the large number of data points (948) makes the difference between these
two quantities irrelevant at the level of precision of the first decimal
place.} and also with respect to the regression on the market portfolio
and either the size or the book-to-market factor in the group ``Small''.
In contrast, for portfolios belonging to the two last quintiles of the
distribution of size, {\it i.e.}, portfolios of the group 4 and Big, the
factor HML provides the largest improvement 9 times out of 10 and is
statistically significant, with respect to the regression on the sole
market factor, for 8 of these portfolios.

For the {\em three-factor} models, the
pair (SMB,HML) provides the best improvement in 13 cases out of 25,
before the pair (ICC,HML) which is the best 8 times out of 25, while the
pair (SMB,ICC) wins the ``horse race'' only 4 times out of 25. However,
these improvement are statically significant with
respect to the best two-factor model (which is most
often the market $+$ factor ICC) in only 5 cases out of 25, namely for the
portfolios 2-4, 2-High; 3-4, 3-High; and 4-Low. Therefore, the usefulness of a three-factor
model is clearly questionable. 

To sum up our tests performed on the 25 equally-weighted portfolios
ordered by quintiles in size and book-to-market, we have found that, on
average, the factor ICC alone provides the best significant improvement
with respect to the market factor, and also provides a significant
improvement with respect to the market factor and either the size or the
book-to-market factor. Overall, the addition of one or two of the Fama
and French factors turns out to provide only a marginal improvement.
The confidence intervals on the $R^2$ obtained by bootstrap suggests
that a two-factor model (market portfolio $+$ factor ICC) has almost the
same explanatory power than the three-factor Fama-French model, while
being more parsimonious and based on solid economic foundation. Beside,
the significance of the intercepts $\alpha$'s remains comparable (see
the last two lines of Table~\ref{tab1b}). In all cases, the GRS test
\cite{GRS89} underlines that the intercept is significantly different
from zero. In this respect, the factor ICC does not really improve on
the two factors of Fama and French but, clearly, the GRS statistics
reaches its minimum when the size factor is replaced by the ICC factor.
Therefore, based on the results on the ``Small'' group of portfolios, on
the GRS test and on our theoretical approach, we can finally conclude to
the superiority of the factor ICC with respect to the size factor SMB.
On the hand, the explaining power of the book-to-market factor HML seems
undisputable even if it is weakened in the presence of ICC.

\vspace{0.5cm}
\centerline{\it [Insert Tables~\ref{tab2a} and~\ref{tab2b} about here]}
\vspace{0.5cm}

The following tables provide the same statistics for value-weighted and
equally-weighted industry portfolios which confirm the previous
conclusions.
Table~\ref{tab2a} presents the parameter estimates of the multi-linear time series
regression of the excess monthly returns of 10 value-weighed
industry portfolios regressed, as in table~\ref{tab1a}, on the excess
return on the market portfolio, on the two Fama-French factors  and on
the factor ICC. In the presence of the risk factor ICC, the
factor SMB turns out to be not significant for most portfolios (7 cases out of 10).
Conversely, in the presence the factor SMB, the factor ICC has no
explanatory power in only 4 cases out of 10. This clearly confirms that,
overall, ICC is a superior substitute to SMB. For the HML factor, the table
shows that this factor is always significant, even in the
presence of the factor ICC. Again, these observations must 
be tempered by the results of table~\ref{tab2b} which provides the $R^2$
of the various multi-linear times series regressions of the monthly
excess returns of these 10 value-weighed industry portfolios on the same
set of factors as in table~\ref{tab1b}. It is striking to observe that,
on the basis of the $95\%$ confidence intervals obtained by bootstrap, none of the factors
ICC, SMB and HML or any combination thereof, is able to provide a
significant improvement with respect to the regression on the sole
market factor (with the exception of the portfolio ``Others''). Concerning
the factor ICC, this observation is
not a big surprise since it is expected to 
provide a strong explanatory power for well-diversified portfolios. But,
by construction, value-weighted portfolios are not diversified, hence the
lack of explanatory power of the factor ICC. Moreover,
if the number of assets in each industry is large enough, we should
expect that the contribution of the residual risk to the total risk goes
to zero, as it goes to zero for the market portfolio.

\vspace{0.5cm}
\centerline{\it [Insert Tables~\ref{tab3a} and~\ref{tab3b} about here]}
\vspace{0.5cm}

The situation is totally different when one considers the same set of
industry portfolios but constructed on an equally-weighted basis. In
this case, each industry portfolio is ``well-diversified,'' in the sense
that the weight of each asset in a given industry portfolio is
inversely proportional to the number of assets in this portfolio.
Tables~\ref{tab3a} and~\ref{tab3b} summarize the values of the
parameter estimates and of the $R^2$, respectively, of the multi-linear
 time series regressions of the excess monthly returns on 10
equally-weighed industry portfolios regressed, as previously, on the
excess return on the market portfolio, on the two Fama-French factors
and on the factor ICC, on the one hand, and on the same set of factors as in
tables~\ref{tab1b} and~\ref{tab2b}, on the other hand. As in the case of
the 25 equally-weighted portfolios sorted by size and book-to-market, 
the addition of the internal consistency
factor ICC to the market factor provides overall the best improvement in terms
of the $R^2$ of two-factor models. In addition, no three- or four-factor
model provides a statistically significant improvement while the GRS test does not reject the hypothesis of a zero-intercept for the model ``Market factor + ICC factor'' at the $2\%$ level.

This confirms that the two-factor model constructed with the
market portfolio and with the internal consistency factor ICC has overall
the same explanatory power as the three-factor Fama-French model.

\subsection{Relation between the internal consistency factor ICC and the
two Fama and French factors SMB and HML}

As illustrated above, the additional internal consistency factor 
allows us to explain several well-known pricing anomalies,
with a power comparable to the HML $+$ SMB Fama-French factors. 
We now discuss why this can be expected on the basis of our theoretical
results. Specifically, starting from our theoretical framework, 
we address the question of why should the 
two additional factors of Fama and French have an explaining power, that is,
what could be the origins of the size and book-to-market effects.

\paragraph{The size effect.} 

\begin{figure}
\begin{center}
\includegraphics[width=12cm]{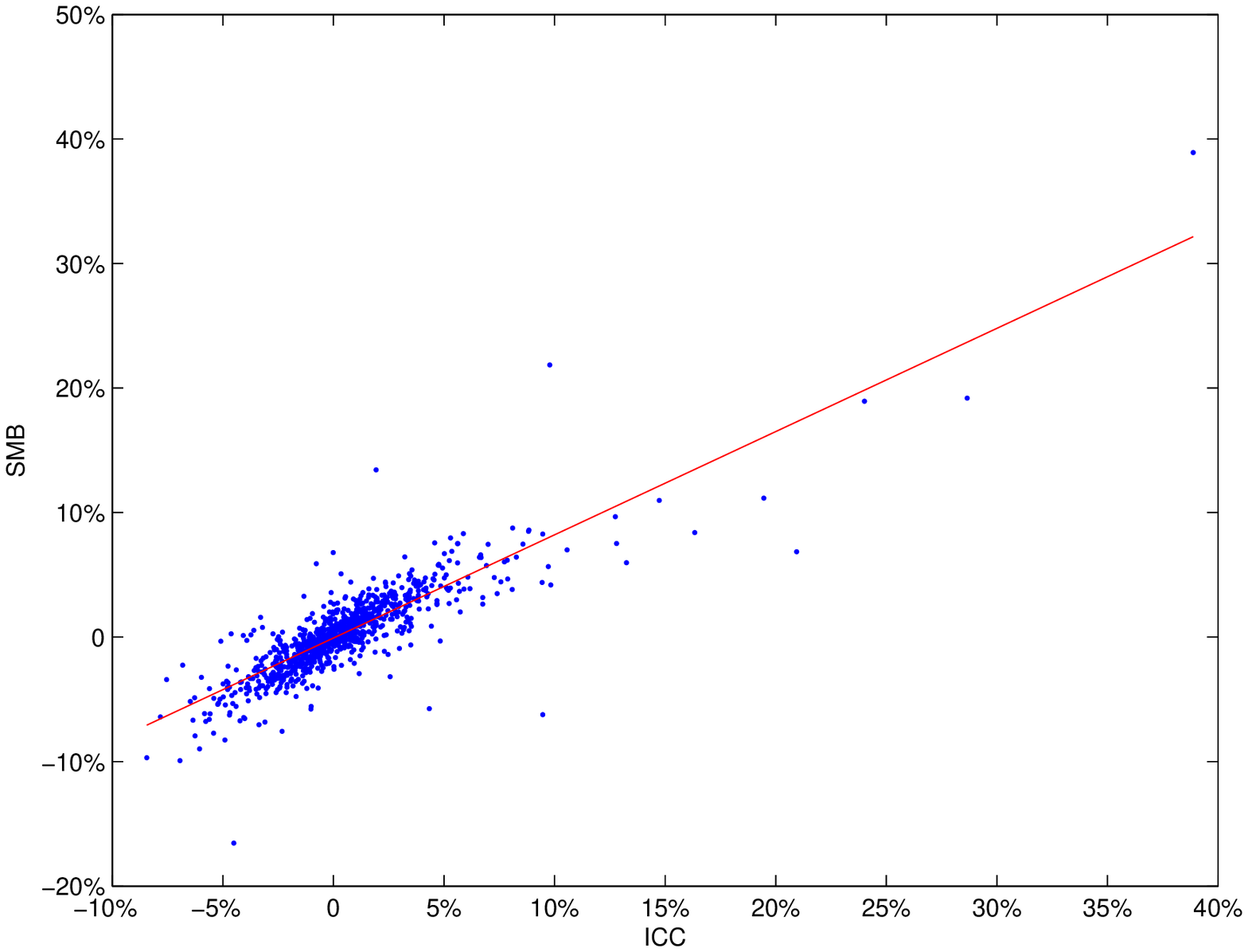}\\
\includegraphics[width=12cm]{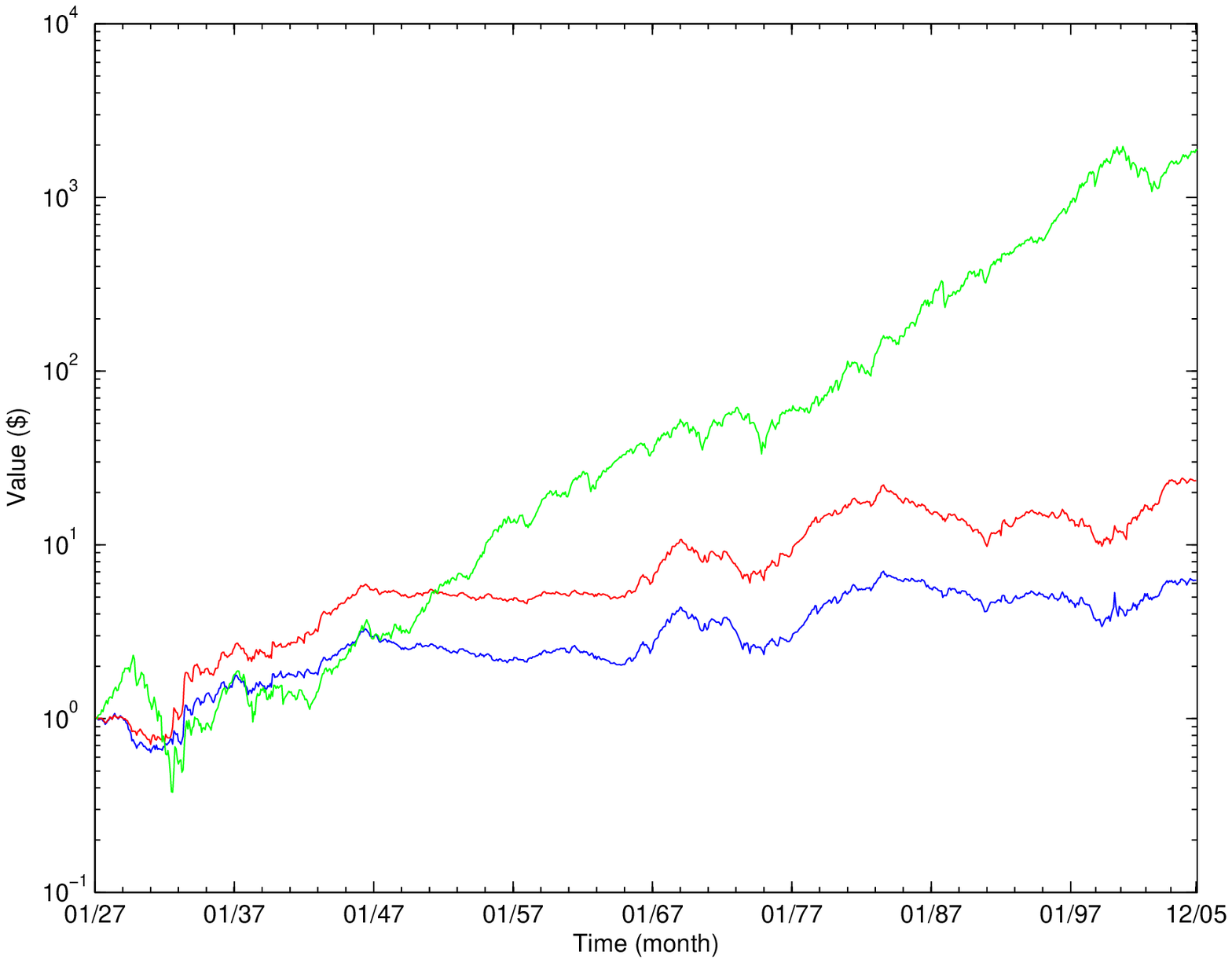}
\end{center}
\captionsetup{style=default,labelfont=bf, labelsep=period}
\caption{\label{fig} {\bf Comparison of ICC and SMB}. The upper panel
shows the return of the factor SMB versus the return of the factor ICC. 
The straight line shows the regression line with
equation $y = -0.0008 + 0.8292 \cdot x$. The lower panel depicts the
value of \$1 invested in the market portfolio in Jan. 1927 (grey curve; green online)
and the value of a leveraged position of \$1 invested in SMB (dark grey curve; blue online) 
and ICC (black curve; red online) in Jan. 1927. For the two arbitrage portfolios SMB and ICC, 
the initial endowment of \$1 can be thought of as a reserve to ensure against risk losses,
from which the returns can be discounted to provide the shown wealth curves.
}
\end{figure}

The size effect is well-known to generally explain the part of the
cross-section of expected returns left unexplained by any misspecified
asset pricing model \cite{Berk95}, which raises the question of its relevance
as the signature of a genuine risk factor.
Our theoretical model provides an answer to this question by
rationalizing the role of the size effect as providing a proxy for the diversification
factor $f$ (or ICC). Indeed,
since the arbitrage portfolio which proxies the ICC factor is long in
the equally-weighted portfolio and short in the market portfolio, it
is therefore long on the small caps and short on the large caps, just like
the SMB portfolio. There is thus no qualitative difference between the
Fama and French's factor SMB and our proxy of the ICC factor. This is
confirmed by the large value of the linear correlation between the two portfolios
proxying the SMB and ICC factors equal to 86\% over the time interval
studied here. As an illustration, the return on each
factors is depicted on the left panel of figure~\ref{fig} while the
right panel represents the value of \$1 invested in the market portfolio
in Jan. 1927 and the value of a leveraged position of \$1
invested in SMB and ICC in Jan. 1927.

\vspace{0.5cm}
\centerline{\it [Insert figure~\ref{fig} about here]}
\vspace{0.5cm}

\paragraph{The book-to-market effect.}

As illustrated by \citeasnoun{Stattman80} and \citeasnoun{RRL85} in the early eighties and as emphasized more recently by \citename{FamaandFrench1993}~\citeyear{FamaandFrench1992,FamaandFrench1993}, stocks with a high book-to-market value
tend to overperform stocks with a low book-to-market value. Several
economic explanations have been proposed to justify this phenomenon.
Among others, \citename{FamaandFrench1996} have proposed that value
stocks are companies that are in financial distress while
\citeasnoun{CV2004} have suggested that growth stocks might have
speculative investment opportunities that will be profitable only if
equity financing is available on sufficiently good terms.

The pricing formula provided by proposition~\ref{prop4} offers a
straightforward justification of the book-to-market effect. Indeed, there is good
empirical evidence that
high book-to-market stocks have significantly lower beta's with respect to the market
portfolio compared with low book-to-market stocks. For instance,
using a large sample of firms from 1977 to 2004, \citeasnoun{Bernardoetal2005} find
that the difference between the beta's of growth opportunities and the beta's of
assets-in-place is positive and statistically significant, at the 95\%
level, in 34 out of 37 industry classifications. Bernardo et al. suggest that this
results from the fact that, since firms with more
growth opportunities have cash flows with longer duration, their values
are more sensitive to changes in interest rates and thus should have
higher beta's. Then, {\it ceteris paribus}, the additional
term $\(\gamma_i - \gamma_m \cdot \beta_i \) \cdot \E\[r_{icc} - r_0\]$ 
introduced by the internal consistency constraint leads to a higher
expected rate of return for a stock with a low beta if the term $\gamma_m$ is positive.

\section{Conclusion}

Starting from a factorial model in which the only a priori systematic
risk is the market portfolio, we have shown that there is a new
source of significant systematic risk, which has been totally neglected
up to now but which ought to be priced by the market. This occurs when
(i) the internal consistency condition holds (which simply means that
the market portfolio is constituted of the assets whose returns it is
supposed to explain) and (ii) the distribution of the capitalization of
firms is sufficiently fat-tailed, as is the
case of real economies. The corresponding new internal consistent factors do not
disappear for arbitrary large economies because the contribution, to the
risk of arbitrary well-diversified portfolios due to the largest firms,
remains finite for arbitrary large economies when the distribution of the
capitalization of firms is sufficiently heavy-tailed. For this reason, this endogenous factor
can be considered as related to the existence of a diversification/concentration 
premium resulting from the concern of investors with respect to
 the level of diversification of their portfolio in so far 
as holding the market portfolio alone does not allow for a good diversification.

Applied to the Arbitrage Pricing Theory, we have shown that the original
derivation of Ross' results still holds, provided that we explicitly
include the additional diversification factor in the
analysis. As a consequence, this factor is
shown to provide possible theoretical economic explanations of some of
the empirical factors reported in the literature. In particular, it 
allows understanding the superior performance of Fama and French
three-factor model in explaining the cross section of stock returns. 
Indeed, the diversification factor provides a rationalization
of the SMB factor as a proxy of this factor. Beside, being
consistent with the fact that high book-to-market stocks have
significantly lower beta's with respect to the market portfolio compared
with low book-to-market stocks, the Value/Growth effect is related to the increasing sensitivity of value stocks to the diversification factor. Finally, on the basis of only two
factors (the market portfolio and the equally-weighted portfolio), our
model turns out to be at least as successful as the Fama and French
three-factor model in explaining the cross-section of monthly returns on US stock over the time period for Jan. 1927 to Dec 2005.

\pagebreak


\clearpage


\clearpage

\begin{landscape}
\begin{table}
\caption{\label{tab:NumSim} Numerical simulations}

Average, minimum and maximum value of the
$R^2$ of the regression of the return of 20 equally weighted portfolios
(randomly drawn from a market of $N=1000$ and $N=10000$ assets according to the model~(\ref{eq:capm2})) on the market
portfolio ($r_m$), on the market portfolio and the internal consistency
factor ($r_m, f$), on the market portfolio and the (overall) equally
weighted portfolio ($r_m, r_e$), on the market portfolio and an
under-diversified portfolio ($r_m, r_u$) and on the market portfolio and
a well-diversified arbitrage portfolios ($r_m, r_a$). Different market
situations are considered with distributions of firm sizes with tail
index $\mu$ which varies from $0.5$ to $2$.
\vspace{0.5cm}

\begin{center}
\begin{tabular}{clrrrrrcrrrrr}
\hline
\hline
 &  & \multicolumn{5}{c}{N=1000} &  & \multicolumn{5}{c}{N=10000} \\
\cline{3-7}
\cline{9-13}
 &  & $r_m$ & $r_m, f$ & $r_m, r_{e}$ & $r_m, r_{u}$ & $r_m, r_{a}$ & & $r_m$ & $r_m, f$ & $r_m, r_{e}$ & $r_m, r_{u}$ & $r_m, r_{a}$\\
\\
 & Mean & 94\% & 94\% & 95\% & 94\% & 94\% &  & 99\% & 99\% & 99\% & 99\% & 99\% \\
$\mu=2$ & Min & 90\% & 93\% & 93\% & 90\% & 90\% &  & 99\% & 99\% & 99\% & 99\% & 99\% \\
 & Max & 96\% & 96\% & 96\% & 96\% & 96\% &  & 100\% & 100\% & 100\% & 100\% & 100\% \\
 &  &  &  &  &  &  &  &  &  &  &  &  \\
 & Mean & 80\% & 95\% & 95\% & 86\% & 82\% &  & 88\% & 99\% & 99\% & 93\% & 89\% \\
$\mu=1$ & Min & 1\% & 91\% & 91\% & 42\% & 17\% &  & 20\% & 99\% & 99\% & 66\% & 20\% \\
 & Max & 95\% & 100\% & 100\% & 95\% & 95\% &  & 99\% & 100\% & 100\% & 99\% & 99\% \\
\\
 & Mean & 56\% & 97\% & 97\% & 79\% & 64\% &  & 56\% & 100\% & 100\% & 83\% & 63\% \\
$\mu=1/2$ & Min & 2\% & 89\% & 89\% & 34\% & 15\% &  & 1\% & 96\% & 97\% & 15\% & 3\% \\
 & Max & 100\% & 100\% & 100\% & 100\% & 100\% &  & 100\% & 100\% & 100\% & 100\% & 100\% \\
\\
\hline
\hline
\end{tabular}
\end{center}
\end{table}
\end{landscape}

\clearpage

\begin{table}
\caption{\label{tab1a} Multi-factor time series regressions for monthly excess returns on
25 equally-weighted portfolios sorted by size and book-to-market (beta):
Jan. 1927 - Dec. 2005, 948 months}

Parameter estimates of the linear regression of
the excess returns on 25 equally-weighed portfolios (sorted by quintiles
of the distribution of size -- Small, 2, 3, 4 and Big -- and by
quintiles of the distribution of Book equity to Market equity ratio --
Low, 2, 3, 4 and High) regressed on
the excess return on the market portfolio, on the two Fama-French
factors SMB and HML and on the proxy for the additional risk factor due
to the internal consistency constraint given by the difference between
the return on the equally-weighted portfolio and the return on the
market portfolio:
$$r_{i,t} - r_0 = \alpha_i + \beta_i \cdot \[r_m(t) -
r_0\] + \beta^{ICC}_{i} \cdot r_{icc}(t) + \beta^{SMB}_{i} \cdot
r_{smb}(t) + \beta^{HML}_{i} \cdot r_{hml}(t) +\epsilon_i(t).$$
In the
four columns labeled $\beta$, $\beta^{SMB}$, $\beta^{HML}$ and $\beta^{ICC}$
the figures decorated by one star (resp. two stars) show the cases which
reject the null hypothesis that the factor under consideration is not
significant in the presence of the others at the 5\% (resp. the 1\%)
level. For instance, for the portfolio Big-High, the factor SMB is
not significant (neither at the 5\% nor the 1\% level) in the presence of
both the market factor, the factor HML and the proxy for the factor ICC.
Similarly, the factor ICC is not significant in the presence of the market factor, the SMB
and HML factors while, in contrast, the factor HML is still significant at
the $1\%$ level in the presence of the market factor, the the SMB and ICC
factors.
\vspace{0.5cm}

\begin{center}
 \begin{tabular}{llcccccc}
\hline
\hline
 &  & $\alpha$ & $\beta$ & $\beta^{SMB}$ & $\beta^{HML}$ & $\beta^{ICC}$ & $R^2$ \\
\hline
 & Low & -0.0076 & 1.24$^{**}$ & -0.49$^{**}$ & -0.24$^{**}$ & 2.33$^{**}$ & 75\% \\
 & 2 & -0.0032 & 1.05$^{**}$ & 0.78$^{**}$ & 0.16$^{*}$ & 1.17$^{**}$ & 81\% \\
Small & 3 & 0.0007 & 1.01$^{**}$ & 0.37$^{**}$ & 0.21$^{**}$ & 1.06$^{**}$ & 89\% \\
 & 4 & 0.0017 & 0.94$^{**}$ & 0.47$^{**}$ & 0.36$^{**}$ & 1.05$^{**}$ & 94\% \\
 & High & 0.0037 & 0.93$^{**}$ & 0.45$^{**}$ & 0.65$^{**}$ & 1.32$^{**}$ & 92\% \\
 &  &  &  &  &  &  &  \\[-2mm]
 & Low & -0.0032 & 1.11$^{**}$ & 0.70$^{**}$ & -0.38$^{**}$ & 0.56$^{**}$ & 90\% \\
 & 2 & -0.0009 & 1.11$^{**}$ & 0.69$^{**}$ & 0.14$^{**}$ & 0.34$^{**}$ & 94\% \\
2 & 3 & 0.0011 & 0.98$^{**}$ & 0.75$^{**}$ & 0.33$^{**}$ & 0.20$^{**}$ & 93\% \\
 & 4 & 0.0008 & 1.00$^{**}$ & 0.74$^{**}$ & 0.56$^{**}$ & 0.11$^{**}$ & 95\% \\
 & High & -0.0004 & 1.07$^{**}$ & 0.79$^{**}$ & 0.83$^{**}$ & 0.19$^{**}$ & 96\% \\
 &  &  &  &  &  &  &  \\[-2mm]
 & Low & -0.0021 & 1.16$^{**}$ & 0.29$^{**}$ & -0.38$^{**}$ & 0.61$^{**}$ & 92\% \\
 & 2 & 0.0010 & 1.03$^{**}$ & 0.44$^{**}$ & 0.03 & 0.11$^{*}$ & 92\% \\
3 & 3 & 0.0005 & 1.04$^{**}$ & 0.38$^{**}$ & 0.32$^{**}$ & 0.08$^{**}$ & 93\% \\
 & 4 & 0.0011 & 0.97$^{**}$ & 0.51$^{**}$ & 0.52$^{**}$ & -0.01$^{**}$ & 93\% \\
 & High & -0.0007 & 1.18$^{**}$ & 0.31$^{**}$ & 0.87$^{**}$ & 0.28$^{**}$ & 94\% \\
 &  &  &  &  &  &  &  \\[-2mm]
 & Low & 0.0004 & 1.08$^{**}$ & 0.07 & -0.44$^{**}$ & 0.26$^{**}$ & 93\% \\
 & 2 & -0.0004 & 1.04$^{**}$ & 0.14$^{**}$ & 0.10$^{**}$ & 0.11$^{*}$ & 91\% \\
4 & 3 & 0.0010 & 1.02$^{**}$ & 0.17$^{**}$ & 0.29$^{**}$ & 0.09 & 92\% \\
 & 4 & 0.0002 & 1.08$^{**}$ & 0.08 & 0.57$^{**}$ & 0.16$^{**}$ & 93\% \\
 & High & -0.0024 & 1.27$^{**}$ & 0.17$^{**}$ & 0.98$^{**}$ & 0.28$^{**}$ & 93\% \\
 &  &  &  &  &  &  &  \\[-2mm]
 & Low & 0.0002 & 1.06$^{**}$ & -0.24$^{**}$ & -0.35$^{**}$ & 0.21$^{**}$ & 96\% \\
 & 2 & 0.0003 & 1.04$^{**}$ & -0.19$^{**}$ & 0.07$^{**}$ & 0.13$^{**}$ & 94\% \\
Big & 3 & -0.0001 & 1.04$^{**}$ & -0.20$^{**}$ & 0.32$^{**}$ & 0.11$^{**}$ & 93\% \\
 & 4 & -0.0015 & 1.10$^{**}$ & -0.30$^{**}$ & 0.66$^{**}$ & 0.26$^{**}$ & 92\% \\
 & High & -0.0012 & 1.10$^{**}$ & -0.26$^{**}$ & 0.82$^{**}$ & 0.27$^{**}$ & 86\% \\
 \hline
\hline
\end{tabular}
\end{center}
\end{table}

\begin{landscape}
\begin{table}
\caption{\label{tab1b} Multi-factor time series regressions for monthly excess returns on
25 equally-weighted portfolios sorted by size and book-to-market ($R^2$):
Jan. 1927 - Dec. 2005, 948 months}

$R^2$ of the linear regression of
the excess returns of 25 equally-weighed portfolios (sorted by quintiles
of the distribution of size -- Small, 2, 3, 4 and Big -- and by
quintiles of the distribution of Book equity to Market equity ratio --
Low, 2, 3, 4 and High) on the market portfolio (Rm), on the market
portfolio and the factor ICC (ICC), on the market portfolio and the size
factor (SMB), on the market portfolio and the book to market factor
(HML), on the market portfolio and the two Fama\&French factors (HML +
SMB),  on the market portfolio, the factor ICC and the size factor (ICC
+ SMB), on the market portfolio, the factor ICC and the book to market
factor (ICC + HML) and, finally on all these four factors (Market, ICC, SMB 
and HML). Figures in
boldface represent the maximum value of the $R^2$ within the group of
regression with two factors (columns ICC, SMB and HML) and with three
factors (columns HML + SMB, ICC + SMB and ICC + HML). The two last rows
reports \citeasnoun{GRS89} test statistics and $p$-values.
\vspace{0.5cm}

\begin{center}
\begin{tabular}[c]{llccccccccccc}
\hline
\hline
 &  &  &  &  &  &  &  & HML & ICC & ICC &  & All  \\[-2mm]
 &  & Rm &  & ICC & SMB & HML &  & + & + & + &  & four  \\[-2mm]
 &  &  &  &  &  &  &  & SMB & SMB & HML &  & factors \\ 
\cline{3-3}
\cline{5-7}
\cline{9-11}
\cline{13-13}
& Low & 52.0\% &  & \textbf{ 74.8\%} & 66.7\% & 54.3\% &  & 68.6\% & \textbf{ 74.9\%} & 74.8\% &  & 75.2\% \\[-2mm]
&& {\tiny (43.1\%,60.6\%)} &&{\tiny (68.5\%,80.3\%)} &{\tiny (60.1\%,74.5\%)} &{\tiny (44.3\%,64.3\%)} &&{\tiny (62.1\%,75.8\%)} &{\tiny (68.7\%,80.9\%)} &{\tiny (69.3\%,80.6\%)} &&{\tiny (69.6\%,81.0\%)}\\[1mm]
 & 2 & 51.8\% &  & \textbf{ 79.9\%} & 76.4\% & 54.9\% &  & 78.9\% & \textbf{ 80.7\%} & 79.9\% &  & 80.9\% \\[-2mm]
&& {\tiny (43.4\%,61.5\%)} &&{\tiny (73.0\%,86.0\%)} &{\tiny (71.9\%,81.3\%)} &{\tiny (45.7\%,66.6\%)} &&{\tiny (73.3\%,84.3\%)} &{\tiny (75.1\%,86.1\%)} &{\tiny (74.0\%,86.2\%)} &&{\tiny (75.5\%,86.5\%)}\\[1mm]
Small  & 3 & 63.8\% &  & \textbf{ 89.0\%} & 82.9\% & 68.5\% &  & 87.0\% & 89.1\% & \textbf{ 89.1\%} &  & 89.4\% \\[-2mm]
&& {\tiny (57.2\%,70.3\%)} &&{\tiny (85.8\%,91.8\%)} &{\tiny (80.0\%,85.6\%)} &{\tiny (60.7\%,76.5\%)} &&{\tiny (83.9\%,90.2\%)} &{\tiny (86.4\%,91.9\%)} &{\tiny (85.8\%,92.5\%)} &&{\tiny (86.6\%,92.6\%)}\\[1mm]
 & 4 & 61.7\% &  & \textbf{ 92.5\%} & 84.4\% & 69.4\% &  & 91.3\% & 92.6\% & \textbf{ 93.2\%} &  & 93.7\% \\[-2mm]
&& {\tiny (53.8\%,69.8\%)} &&{\tiny (90.9\%,94.2\%)} &{\tiny (81.6\%,87.6\%)} &{\tiny (62.1\%,77.2\%)} &&{\tiny (89.3\%,93.2\%)} &{\tiny (91.0\%,94.3\%)} &{\tiny (91.7\%,95.0\%)} &&{\tiny (92.5\%,95.3\%)}\\[1mm]
 & High & 53.9\% &  & \textbf{ 89.5\%} & 77.2\% & 67.5\% &  & 89.6\% & 89.7\% & \textbf{ 92.1\%} &  & 92.5\% \\[-2mm]
&& {\tiny (46.3\%,62.6\%)} &&{\tiny (86.0\%,92.5\%)} &{\tiny (71.2\%,82.5\%)} &{\tiny (60.9\%,74.3\%)} &&{\tiny (85.9\%,92.4\%)} &{\tiny (86.1\%,92.7\%)} &{\tiny (89.5\%,94.4\%)} &&{\tiny (89.9\%,94.8\%)}\\[1mm]
 &  &  &  &  &  &  &  &  &  &  &  &  \\
 & Low & 70.3\% &  & 84.2\% & \textbf{ 88.9\%} & 70.8\% &  & \textbf{ 89.6\%} & 88.9\% & 89.0\% &  & 90.4\% \\[-2mm]
&& {\tiny (66.1\%,75.4\%)} &&{\tiny (81.0\%,87.7\%)} &{\tiny (86.1\%,91.5\%)} &{\tiny (66.5\%,76.3\%)} &&{\tiny (87.1\%,92.1\%)} &{\tiny (86.4\%,91.5\%)} &{\tiny (86.7\%,91.5\%)} &&{\tiny (88.5\%,92.6\%)}\\[1mm]
 & 2 & 78.0\% &  & 92.2\% & \textbf{ 92.3\%} & 79.3\% &  & 93.4\% & \textbf{ 93.5\%} & 92.3\% &  & 93.7\% \\[-2mm]
&& {\tiny (71.3\%,84.1\%)} &&{\tiny (90.3\%,94.1\%)} &{\tiny (90.8\%,94.0\%)} &{\tiny (73.2\%,85.0\%)} &&{\tiny (92.1\%,94.9\%)} &{\tiny (92.3\%,95.0\%)} &{\tiny (90.5\%,94.2\%)} &&{\tiny (92.5\%,95.2\%)}\\[1mm]
2 & 3 & 74.6\% &  & \textbf{ 90.8\%} & 89.6\% & 78.4\% &  & \textbf{ 92.9\%} & 91.6\% & 91.1\% &  & 93.0\% \\[-2mm]
&& {\tiny (65.9\%,83.0\%)} &&{\tiny (88.3\%,93.8\%)} &{\tiny (86.9\%,92.8\%)} &{\tiny (71.0\%,85.9\%)} &&{\tiny (91.2\%,95.3\%)} &{\tiny (89.5\%,94.2\%)} &{\tiny (88.8\%,94.1\%)} &&{\tiny (91.4\%,95.4\%)}\\[1mm]
 & 4 & 75.8\% &  & \textbf{ 91.0\%} & 87.7\% & 83.6\% &  & \textbf{ 94.9\%} & 91.1\% & 93.2\% &  & 95.0\% \\[-2mm]
&& {\tiny (69.2\%,81.8\%)} &&{\tiny (88.4\%,93.1\%)} &{\tiny (84.7\%,90.7\%)} &{\tiny (78.5\%,88.5\%)} &&{\tiny (93.7\%,96.2\%)} &{\tiny (88.6\%,93.2\%)} &{\tiny (91.4\%,94.8\%)} &&{\tiny (93.8\%,96.2\%)}\\[1mm]
 & High & 71.3\% &  & \textbf{ 89.3\%} & 83.4\% & 84.4\% &  & \textbf{ 95.8\%} & 89.4\% & 94.3\% &  & 95.9\% \\[-2mm]
&& {\tiny (65.5\%,76.6\%)} &&{\tiny (85.7\%,92.0\%)} &{\tiny (79.3\%,87.5\%)} &{\tiny (80.4\%,88.0\%)} &&{\tiny (94.0\%,97.0\%)} &{\tiny (85.8\%,92.0\%)} &{\tiny (92.2\%,95.8\%)} &&{\tiny (94.2\%,97.1\%)}\\[1mm]
\hline
\hline
&&&&&&&&&&&& (continued)
\end{tabular}
\end{center}
\end{table}

\begin{table}\ContinuedFloat
\captionsetup{style=default,labelfont=bf,labelsep=endash}
\caption{Continued}
\begin{center}
\begin{tabular}[c]{llccccccccccc}
\hline
\hline
 &  &  &  &  &  &  &  & HML & ICC & ICC &  & All  \\[-2mm]
 &  & Rm &  & ICC & SMB & HML &  & + & + & + &  & four  \\[-2mm]
 &  &  &  &  &  &  &  & SMB & SMB & HML &  & factors \\ 
\cline{3-3}
\cline{5-7}
\cline{9-11}
\cline{13-13}
\\[-3mm] 
 & Low & 80.3\% &  & 88.6\% & \textbf{ 90.7\%} & 80.8\% &  & 91.4\% & 90.8\% & \textbf{ 92.2\%} &  & 92.5\% \\[-2mm]
&& {\tiny (75.7\%,84.8\%)} &&{\tiny (86.1\%,90.8\%)} &{\tiny (87.8\%,93.0\%)} &{\tiny (76.5\%,85.5\%)} &&{\tiny (88.9\%,93.5\%)} &{\tiny (88.4\%,93.1\%)} &{\tiny (90.6\%,93.7\%)} &&{\tiny (90.9\%,94.0\%)}\\[1mm]
 & 2 & 85.6\% &  & 90.9\% & \textbf{ 91.8\%} & 85.7\% &  & 92.0\% & \textbf{ 92.0\%} & 91.1\% &  & 92.0\% \\[-2mm]
&& {\tiny (82.7\%,88.3\%)} &&{\tiny (89.1\%,92.9\%)} &{\tiny (89.7\%,93.8\%)} &{\tiny (82.9\%,88.6\%)} &&{\tiny (90.1\%,93.9\%)} &{\tiny (90.1\%,93.9\%)} &{\tiny (89.2\%,93.0\%)} &&{\tiny (90.2\%,93.9\%)}\\[1mm]
3 & 3 & 85.4\% &  & \textbf{ 91.4\%} & 89.9\% & 88.8\% &  & \textbf{ 93.0\%} & 91.4\% & 92.4\% &  & 93.1\% \\[-2mm]
&& {\tiny (81.9\%,88.4\%)} &&{\tiny (89.2\%,93.2\%)} &{\tiny (87.3\%,92.1\%)} &{\tiny (86.3\%,90.9\%)} &&{\tiny (91.5\%,94.3\%)} &{\tiny (89.3\%,93.3\%)} &{\tiny (90.8\%,93.8\%)} &&{\tiny (91.5\%,94.3\%)}\\[1mm]
 & 4 & 80.4\% &  & \textbf{ 88.7\%} & 86.0\% & 87.8\% &  & \textbf{ 93.0\%} & 88.7\% & 91.9\% &  & 93.0\% \\[-2mm]
& & {\tiny (75.2\%,84.9\%)} &&{\tiny (85.0\%,91.6\%)} &{\tiny (82.2\%,89.4\%)} &{\tiny (84.3\%,91.1\%)} &&{\tiny (91.1\%,94.7\%)} &{\tiny (85.2\%,91.7\%)} &{\tiny (89.6\%,93.9\%)} &&{\tiny (91.1\%,94.7\%)}\\[1mm]
 & High & 75.6\% &  & 85.9\% & 79.9\% & \textbf{ 90.5\%} &  & \textbf{ 94.3\%} & 87.2\% & 94.2\% &  & 94.4\% \\[-2mm]
& & {\tiny (70.8\%,79.7\%)} &&{\tiny (82.5\%,88.9\%)} &{\tiny (75.8\%,83.9\%)} &{\tiny (87.1\%,93.1\%)} &&{\tiny (92.4\%,95.8\%)} &{\tiny (83.7\%,90.3\%)} &{\tiny (92.0\%,95.8\%)} &&{\tiny (92.6\%,96.0\%)}\\[1mm]
 &  &  &  &  &  &  &  &  &  &  &  &  \\[-2mm]
 & Low & 86.4\% &  & 87.0\% & 88.4\% & \textbf{ 90.2\%} &  & 92.3\% & 89.0\% & \textbf{ 92.6\%} &  & 92.6\% \\[-2mm]
&& {\tiny (84.0\%,88.7\%)} &&{\tiny (84.8\%,89.3\%)} &{\tiny (86.2\%,90.4\%)} &{\tiny (88.3\%,91.8\%)} &&{\tiny (90.8\%,93.7\%)} &{\tiny (86.9\%,91.2\%)} &{\tiny (91.4\%,93.9\%)} &&{\tiny (91.4\%,94.0\%)}\\[1mm]
 & 2 & 89.4\% &  & \textbf{ 91.3\%} & 90.8\% & 90.0\% &  & \textbf{ 91.4\%} & 91.3\% & 91.4\% &  & 91.5\% \\[-2mm]
& & {\tiny (87.1\%,91.5\%)} &&{\tiny (89.0\%,93.3\%)} &{\tiny (88.2\%,93.1\%)} &{\tiny (88.0\%,91.9\%)} &&{\tiny (89.3\%,93.5\%)} &{\tiny (89.1\%,93.4\%)} &{\tiny (89.3\%,93.4\%)} &&{\tiny (89.4\%,93.5\%)}\\[1mm]
4 & 3 & 87.3\% &  & 90.3\% & 88.9\% & \textbf{ 90.5\%} &  & \textbf{ 92.0\%} & 90.5\% & 91.9\% &  & 92.0\% \\[-2mm]
& & {\tiny (84.5\%,89.8\%)} &&{\tiny (87.5\%,92.5\%)} &{\tiny (86.2\%,91.5\%)} &{\tiny (88.5\%,92.6\%)} &&{\tiny (89.8\%,94.0\%)} &{\tiny (87.8\%,92.7\%)} &{\tiny (89.7\%,93.8\%)} &&{\tiny (89.8\%,94.0\%)}\\[1mm]
 & 4 & 82.5\% &  & 86.6\% & 83.5\% & \textbf{ 91.8\%} &  & 92.7\% & 88.1\% & \textbf{ 92.8\%} &  & 92.8\% \\[-2mm]
& & {\tiny (78.6\%,85.7\%)} &&{\tiny (82.6\%,89.8\%)} &{\tiny (79.9\%,87.1\%)} &{\tiny (89.2\%,93.9\%)} &&{\tiny (90.3\%,94.6\%)} &{\tiny (84.1\%,91.1\%)} &{\tiny (90.2\%,94.6\%)} &&{\tiny (90.3\%,94.7\%)}\\[1mm]
 & High & 74.4\% &  & 82.1\% & 76.6\% & \textbf{ 90.7\%} &  & 92.5\% & 84.5\% & \textbf{ 92.6\%} &  & 92.7\% \\[-2mm]
& & {\tiny (69.6\%,78.9\%)} &&{\tiny (77.6\%,85.9\%)} &{\tiny (72.0\%,81.1\%)} &{\tiny (87.6\%,93.2\%)} &&{\tiny (89.9\%,94.5\%)} &{\tiny (79.7\%,88.7\%)} &{\tiny (90.0\%,94.6\%)} &&{\tiny (90.1\%,94.6\%)}\\[1mm]
 &  &  &  &  &  &  &  &  &  &  &  &  \\[-2mm]
 & Low & 92.0\% &  & 92.5\% & 92.2\% & \textbf{ 95.1\%} &  & \textbf{ 95.2\%} & 92.7\% & 95.1\% &  & 95.5\% \\[-2mm]
&& {\tiny (90.5\%,93.3\%)} &&{\tiny (91.0\%,93.8\%)} &{\tiny (90.7\%,93.5\%)} &{\tiny (94.0\%,96.1\%)} &&{\tiny (94.2\%,96.2\%)} &{\tiny (91.1\%,94.1\%)} &{\tiny (94.0\%,96.1\%)} &&{\tiny (94.6\%,96.4\%)}\\[1mm]
 & 2 & 93.3\% &  & 93.3\% & 93.5\% & \textbf{ 93.7\%} &  & \textbf{ 93.9\%} & 93.9\% & 93.7\% &  & 94.0\% \\[-2mm]
& & {\tiny (91.0\%,94.9\%)} &&{\tiny (91.0\%,95.0\%)} &{\tiny (91.5\%,95.0\%)} &{\tiny (91.6\%,95.3\%)} &&{\tiny (92.0\%,95.4\%)} &{\tiny (92.1\%,95.4\%)} &{\tiny (91.8\%,95.3\%)} &&{\tiny (92.2\%,95.5\%)}\\[1mm]
Big & 3 & 88.2\% &  & 88.3\% & 88.4\% & \textbf{ 92.3\%} &  & \textbf{ 92.7\%} & 90.6\% & 92.5\% &  & 92.7\% \\[-2mm]
& & {\tiny (85.0\%,90.6\%)} &&{\tiny (85.1\%,90.9\%)} &{\tiny (85.6\%,90.9\%)} &{\tiny (90.0\%,94.2\%)} &&{\tiny (90.5\%,94.5\%)} &{\tiny (87.9\%,93.0\%)} &{\tiny (90.3\%,94.3\%)} &&{\tiny (90.5\%,94.6\%)}\\[1mm]
 & 4 & 79.0\% &  & 80.5\% & 79.1\% & \textbf{ 91.9\%} &  & \textbf{ 92.0\%} & 86.0\% & 91.9\% &  & 92.2\% \\[-2mm]
& & {\tiny (74.3\%,82.9\%)} &&{\tiny (75.8\%,84.5\%)} &{\tiny (74.7\%,83.0\%)} &{\tiny (89.2\%,94.0\%)} &&{\tiny (89.3\%,94.1\%)} &{\tiny (81.5\%,89.6\%)} &{\tiny (89.2\%,94.0\%)} &&{\tiny (89.6\%,94.3\%)}\\[1mm]
 & High & 70.1\% &  & 72.6\% & 70.1\% & \textbf{ 86.2\%} &  & 86.2\% & 78.5\% & \textbf{ 86.3\%} &  & 86.5\% \\[-2mm]
& & {\tiny (64.1\%,75.2\%)} &&{\tiny (66.9\%,77.2\%)} &{\tiny (64.4\%,75.3\%)} &{\tiny (82.4\%,89.9\%)} &&{\tiny (82.5\%,89.9\%)} &{\tiny (72.7\%,83.3\%)} &{\tiny (82.5\%,89.9\%)} &&{\tiny (82.8\%,90.1\%)}\\
 &  &  &  &  &  &  &  &  &  &  &  &  \\[-2mm]
\multicolumn{2}{c}{Average} & 76.1\% &  & \textbf{ 87.3\%} & 84.7\% & 82.2\% &  & 90.6\% & 88.6\% & \textbf{ 90.8\%} &  & 91.4\% \\[-2mm]
&& {\tiny (72.6\%,79.7\%)} &&{\tiny (85.3\%,89.3\%)} &{\tiny (82.4\%,87.2\%)} &{\tiny (79.4\%,85.3\%)} &&{\tiny (89.1\%,92.2\%)} &{\tiny (86.7\%,90.6\%)} &{\tiny (89.5\%,92.2\%)} &&{\tiny (90.2\%,92.8\%)}\\
\\
\multicolumn{2}{c}{GRS} & 4.37 &  & 4.11 & 4.41 & 4.02 &  & 4.07 & 4.19 & 3.92 &  & 4.06 \\
\multicolumn{2}{c}{p-value} & 0.00 &  & 0.00 & 0.00 & 0.00 &  & 0.00 & 0.00 & 0.00 &  & 0.00 \\
\hline
\hline
\end{tabular}
\end{center}
\end{table}
\end{landscape}

\clearpage

\begin{table}
\caption{\label{tab2a} Multi-factor time series regressions for monthly excess returns on
10 value-weighted industry portfolios (beta): Jan. 1927 - Dec. 2005, 948 months}

 Parameter estimates of the linear regression of
the excess returns of ten value-weighed industry portfolios regressed on
the excess return on the market portfolio, on the two Fama-French
factors SMB and HML and on the proxy for the additional risk factor due
to the internal consistency constraint given by the difference between
the return on the equally-weighted portfolio and the return on the
market portfolio:
$$r_{i,t} - r_0 = \alpha_i + \beta_i \cdot \[r_m(t) -
r_0\] + \beta^{ICC}_{i} \cdot r_{icc}(t) + \beta^{SMB}_{i} \cdot
r_{smb}(t) + \beta^{HML}_{i} \cdot r_{hml}(t) +\epsilon_i(t).$$
In the
four columns labeled $\beta$, $\beta^{SMB}$, $\beta^{HML}$ and $\beta^{ICC}$,
the figures decorated by one star (resp. two stars) show the cases which
reject the null hypothesis that the factor under consideration is not
significant in the presence of the others at the 5\% (resp. the 1\%)
level. For instance, for the Shops, the Health and the Utilities industries, the factor SMB is
not significant (neither at the 5\% nor the 1\% level) in the presence of
both the market factor, the factor HML and the proxy for the factor ICC.
Similarly, for the same industry portfolio, the factor ICC is not
significant in the presence of the market factor and the SMB
and HML factors.
\vspace{0.5cm}

\begin{center}
\begin{tabular}{lcccccc}
\hline
\hline
Industry & $\alpha$ & $\beta$ & $\beta^{SMB}$ & $\beta^{HML}$ & $\beta^{ICC}$ & $R^2$ \\
 \hline
Consumer Non Durables  & 0.0019 & 0.78$^{**}$ & 0.08 & 0.07$^{*}$ & -0.15$^{**}$ & 78\% \\
Consumer Durables   & -0.0006 & 1.11$^{**}$ & -0.12 & 0.11$^{*}$ & 0.24$^{**}$ & 75\% \\
Manufacturing   & -0.0006 & 1.10$^{**}$ & 0.11$^{**}$ & 0.19$^{**}$ & -0.11$^{*}$ & 92\% \\
Energy   & 0.0018 & 0.86$^{**}$ & -0.10 & 0.30$^{**}$ & -0.16 & 64\% \\
Business Equipment   & 0.0012 & 1.27$^{**}$ & -0.21$^{**}$ & -0.45$^{**}$ & 0.37$^{**}$ & 84\% \\
Telecom   & 0.0015 & 0.69$^{**}$ & -0.27$^{**}$ & -0.15$^{**}$ & 0.18$^{**}$ & 63\% \\
Shops   & 0.0010 & 0.96$^{**}$ & 0.03 & -0.14$^{**}$ & 0.06 & 80\% \\
Health   & 0.0030 & 0.91$^{**}$ & -0.01 & -0.15$^{**}$ & -0.11 & 68\% \\
Utilities   & -0.0001 & 0.79$^{**}$ & -0.05 & 0.35$^{**}$ & -0.11 & 63\% \\
Others   & -0.0016 & 1.06$^{**}$ & -0.07 & 0.30$^{**}$ & 0.12$^{**}$ & 92\% \\
\hline
\hline
\end{tabular}
\end{center}
\end{table}

\clearpage

\begin{landscape}
\begin{table}

\caption{\label{tab2b} Multi-factor time series regressions for monthly excess returns on
10 value-weighted industry portfolios ($R^2$): Jan. 1927 - Dec. 2005, 948 months}

 $R^2$ of the linear regressions of
the excess returns of ten value-weighed industry portfolios regressed on
the (excess return) on the market portfolio (Rm), on the market
portfolio and the factor ICC (ICC), on the market portfolio and the size
factor (SMB), on the market portfolio and the book to market factor
(HML), on the market portfolio and the two Fama\&French factors (HML +
SMB),  on the market portfolio, the factor ICC and the size factor (ICC
+ SMB), on the market portfolio, the factor ICC and the book to market
factor (ICC + HML) and, finally on the four factors (market, ICC, SMB, HML). Figures in
boldface represent the maximum value of the $R^2$ within the group of
regressions with two factors (columns ICC, SMB and HML) and with three
factors (columns HML + SMB, ICC + SMB and ICC + HML). The two last rows
reports \citeasnoun{GRS89} test statistics and $p$-values.
\vspace{0.5cm}

\begin{center}
\begin{tabular}{lccccccccccc}
\hline
\hline
  &  &  &  &  &  &  & HML & ICC & ICC &  & All  \\[-2mm]
  & Rm &  & ICC & SMB & HML &  & + & + & + &  & four  \\[-2mm]
  &  &  &  &  &  &  & SMB & SMB & HML &  & factors \\ 
\cline{2-2}
\cline{4-6}
\cline{8-10}
\cline{12-12}
\multirow{2}{4cm}{Consumer Non Durables} & 77.4\% &  & \textbf{ 77.6\%} & 77.5\% & 77.5\% &  & 77.5\% & 77.6\% & \textbf{ 77.7\%} & \textbf{ } & 77.7\% \\[-2mm]
& {\tiny (72.6\%,81.4\%)} &&{\tiny (72.8\%,81.7\%)} &{\tiny (72.7\%,81.5\%)} &{\tiny (72.8\%,81.6\%)} &&{\tiny (73.0\%,81.7\%)} &{\tiny (72.9\%,81.7\%)} &{\tiny (73.0\%,81.7\%)} &&{\tiny (73.1\%,81.9\%)}\\[1mm]
\multirow{2}{4cm}{Consumer Durables} & 74.0\% &  & 74.6\% & 74.2\% & \textbf{ 74.8\%} & \textbf{ } & 74.9\% & 75.0\% & \textbf{ 75.1\%} & \textbf{ } & 75.1\% \\[-2mm]
& {\tiny (69.0\%,78.4\%)} &&{\tiny (69.8\%,78.9\%)} &{\tiny (69.2\%,78.5\%)} &{\tiny (69.9\%,79.2\%)} &&{\tiny (70.1\%,79.3\%)} &{\tiny (70.2\%,79.4\%)} &{\tiny (70.3\%,79.4\%)} &&{\tiny (70.4\%,79.5\%)}\\[1mm]
\multirow{2}{4cm}{Manufacturing} & 91.6\% &  & 91.7\% & 91.6\% & \textbf{ 92.3\%} & \textbf{ } & \textbf{ 92.3\%} & 91.7\% & 92.3\% &  & 92.4\% \\[-2mm]
& {\tiny (89.6\%,93.2\%)} &&{\tiny (89.7\%,93.4\%)} &{\tiny (89.6\%,93.3\%)} &{\tiny (90.5\%,93.8\%)} &&{\tiny (90.5\%,93.8\%)} &{\tiny (89.8\%,93.5\%)} &{\tiny (90.5\%,93.8\%)} &&{\tiny (90.6\%,93.9\%)}\\[1mm]
\multirow{2}{4cm}{Energy} & 60.1\% &  & 60.4\% & 61.4\% & \textbf{ 62.2\%} & \textbf{ } & 63.6\% & 61.9\% & \textbf{ 63.7\%} & \textbf{ } & 63.7\% \\[-2mm]
& {\tiny (53.9\%,65.3\%)} &&{\tiny (54.2\%,65.8\%)} &{\tiny (55.5\%,66.7\%)} &{\tiny (56.0\%,67.2\%)} &&{\tiny (58.2\%,68.6\%)} &{\tiny (56.0\%,67.2\%)} &{\tiny (58.2\%,68.6\%)} &&{\tiny (58.3\%,68.7\%)}\\[1mm]
\multirow{2}{4cm}{Business Equipment} & 81.3\% &  & 81.3\% & 81.4\% & \textbf{ 83.6\%} & \textbf{ } & 83.8\% & 81.6\% & \textbf{ 84.0\%} & \textbf{ } & 84.2\% \\[-2mm]
& {\tiny (77.3\%,84.6\%)} &&{\tiny (77.5\%,84.6\%)} &{\tiny (77.9\%,84.7\%)} &{\tiny (80.7\%,86.2\%)} &&{\tiny (80.9\%,86.4\%)} &{\tiny (78.1\%,84.9\%)} &{\tiny (81.3\%,86.6\%)} &&{\tiny (81.6\%,86.8\%)}\\[1mm]
\multirow{2}{4cm}{Telecom} & 61.4\% &  & 62.0\% & \textbf{ 62.2\%} & 62.0\% &  & \textbf{ 62.7\%} & 62.2\% & 62.3\% &  & 63.0\% \\[-2mm]
& {\tiny (56.0\%,66.5\%)} &&{\tiny (57.2\%,67.0\%)} &{\tiny (57.3\%,67.2\%)} &{\tiny (56.9\%,67.0\%)} &&{\tiny (58.1\%,67.6\%)} &{\tiny (57.5\%,67.3\%)} &{\tiny (57.6\%,67.2\%)} &&{\tiny (58.5\%,67.9\%)}\\[1mm]
\multirow{2}{4cm}{Shops} & 78.9\% &  & 78.9\% & 79.1\% & \textbf{ 79.4\%} & \textbf{ } & 79.5\% & 79.2\% & \textbf{ 79.6\%} & \textbf{ } & 79.6\% \\[-2mm]
& {\tiny (74.5\%,82.7\%)} &&{\tiny (74.6\%,82.7\%)} &{\tiny (74.8\%,83.1\%)} &{\tiny (75.3\%,83.1\%)} &&{\tiny (75.5\%,83.4\%)} &{\tiny (75.0\%,83.2\%)} &{\tiny (75.5\%,83.3\%)} &&{\tiny (75.6\%,83.4\%)}\\[1mm]
\multirow{2}{4cm}{Health} & 66.0\% &  & 67.0\% & 66.4\% & \textbf{ 67.3\%} & \textbf{ } & 67.7\% & 67.2\% & \textbf{ 67.7\%} & \textbf{ } & 67.7\% \\[-2mm]
& {\tiny (59.1\%,71.3\%)} &&{\tiny (60.8\%,72.4\%)} &{\tiny (59.8\%,71.7\%)} &{\tiny (61.0\%,72.5\%)} &&{\tiny (61.6\%,72.9\%)} &{\tiny (61.5\%,72.7\%)} &{\tiny (61.8\%,73.1\%)} &&{\tiny (62.2\%,73.2\%)}\\[1mm]
\multirow{2}{4cm}{Utilities} & 58.5\% &  & 58.5\% & 59.1\% & \textbf{ 62.2\%} & \textbf{ } & 62.9\% & 60.1\% & \textbf{ 62.9\%} & \textbf{ } & 62.9\% \\[-2mm]
& {\tiny (51.0\%,65.3\%)} &&{\tiny (51.0\%,65.5\%)} &{\tiny (51.8\%,65.6\%)} &{\tiny (55.3\%,68.7\%)} &&{\tiny (56.1\%,69.1\%)} &{\tiny (52.9\%,66.8\%)} &{\tiny (56.4\%,69.2\%)} &&{\tiny (56.5\%,69.2\%)}\\[1mm]
\multirow{2}{4cm}{Others} & 88.4\% &  & 89.2\% & 88.4\% & \textbf{ 91.7\%} & \textbf{ } & 91.7\% & 90.2\% & \textbf{ 91.8\%} & \textbf{ } & 91.8\% \\[-2mm]
& {\tiny (86.5\%,90.2\%)} &&{\tiny (87.2\%,91.1\%)} &{\tiny (86.5\%,90.4\%)} &{\tiny (89.9\%,93.6\%)} &&{\tiny (89.9\%,93.6\%)} &{\tiny (88.4\%,92.2\%)} &{\tiny (90.0\%,93.6\%)} &&{\tiny (90.0\%,93.7\%)}\\[1mm]
 &  &  &  &  &  &  &  &  &  &  &  \\
\multirow{2}{4cm}{Average} & 73.7\% &  & 74.1\% & 74.1\% & \textbf{ 75.2\%} & \textbf{ } & 75.6\% & 74.6\% & \textbf{ 75.6\%} & \textbf{ } & 75.7\% \\[-2mm]
& {\tiny (70.2\%,76.8\%)} &&{\tiny (70.8\%,77.2\%)} &{\tiny (70.8\%,77.2\%)} &{\tiny (72.1\%,78.2\%)} &&{\tiny (72.7\%,78.6\%)} &{\tiny (71.4\%,77.8\%)} &{\tiny (72.6\%,78.6\%)} &&{\tiny (72.9\%,78.7\%)}\\
\\
GRS&	2.67 &&		3.23& 	3.16& 	3.32 &&		3.69& 	3.09& 	3.63 &&		3.65 \\
$p$-value&	0.00 &&		0.00& 	0.00& 	0.00 &&		0.00& 	0.00& 	0.00 &&		0.00 \\
\hline
\hline
\end{tabular}
\end{center}
\end{table}
\end{landscape}

\clearpage

\begin{table}

\caption{\label{tab3a} Multi-factor time series regressions for monthly excess returns on
10 equally-weighted industry portfolios (beta): Jan. 1927 - Dec. 2005, 948 months}

 Parameter estimates of the linear regression of
the excess returns of ten equally-weighed industry portfolios regressed on
the excess return on the market portfolio, on the two Fama-French
factors SMB and HML and on the proxy for the additional risk factor due
to the internal consistency constraint given by the difference between
the return on the equally-weighted portfolio and the return on the
market portfolio:
$$r_{i,t} - r_0 = \alpha_i + \beta_i \cdot \[r_m(t) -
r_0\] + \beta^{ICC}_{i} \cdot r_{icc}(t) + \beta^{SMB}_{i} \cdot
r_{smb}(t) + \beta^{HML}_{i} \cdot r_{hml}(t) +\epsilon_i(t).$$
In the four columns labeled $\beta$, $\beta^{SMB}$, $\beta^{HML}$ and $\beta^{ICC}$,
the figures decorated by one star (resp. two stars) show the cases which
reject the null hypothesis that the factor under consideration is not
significant in the presence of the others at the 5\% (resp. the 1\%)
level.
\vspace{0.5cm}

\begin{center}
\begin{tabular}{lcccccc}
\hline
\hline
Industry & $\alpha$ & $\beta$ & $\beta^{SMB}$ & $\beta^{HML}$ & $\beta^{ICC}$ & $R^2$ \\
 \hline
Consumer Non Durables  & -0.0003 & 0.84$^{**}$ & 0.08$^{*}$ & 0.10$^{**}$ & 0.77$^{**}$ & 94\% \\
Consumer Durables  & -0.0024 & 1.12$^{**}$ & 0.21$^{**}$ & 0.07$^{*}$ & 0.97$^{**}$ & 92\% \\
Manufacturing  & -0.0004 & 1.07$^{**}$ & 0.12$^{**}$ & 0.17$^{**}$ & 0.76$^{**}$ & 97\% \\
Energy  & 0.0019 & 0.95$^{**}$ & 0.13 & 0.34$^{**}$ & 0.55$^{**}$ & 69\% \\
Business Equipment  & 0.0016 & 1.22$^{**}$ & -0.29$^{**}$ & -0.65$^{**}$ & 1.52$^{**}$ & 92\% \\
Telecom  & 0.0030 & 0.92$^{**}$ & -0.30$^{**}$ & -0.54$^{**}$ & 0.98$^{**}$ & 73\% \\
Shops  & 0.0000 & 0.91$^{**}$ & 0.11$^{*}$ & -0.11$^{**}$ & 0.93$^{**}$ & 90\% \\
Health  & 0.0037 & 0.91$^{**}$ & -0.04 & -0.54$^{**}$ & 0.92$^{**}$ & 80\% \\
Utilities  & 0.0006 & 0.85$^{**}$ & 0.21$^{*}$ & 0.55$^{**}$ & -0.06 & 66\% \\
Others  & -0.0008 & 0.95$^{**}$ & 0.07 & 0.39$^{**}$ & 0.93$^{**}$ & 95\% \\\hline
\hline
\end{tabular}
\end{center}
\end{table}

\clearpage

\begin{landscape}

\begin{table}
\caption{\label{tab3b} Multi-factor time series regressions for monthly excess returns on
10 equally-weighted industry portfolios ($R^2$): Jan. 1927 - Dec. 2005, 948 months}

 $R^2$ of the linear regression of
the excess returns of ten equally-weighed industry portfolios regressed
on the (excess return) on the market portfolio (Rm), on the market
portfolio and the factor ICC (ICC), on the market portfolio and the size
factor (SMB), on the market portfolio and the book to market factor
(HML), on the market portfolio and the two Fama\&French factors (HML +
SMB),  on the market portfolio, the factor ICC and the size factor (ICC
+ SMB), on the market portfolio, the factor ICC and the book to market
factor (ICC + HML) and, finally on the four factors (market, ICC, SMB and HML). Figures in
boldface represent the maximum value of the $R^2$ within the group of
regression with two factors (columns ICC, SMB and HML) and with three
factors (columns HML + SMB, ICC + SMB and ICC + HML). The two last rows reports \citeasnoun{GRS89} test statistics and $p$-values.
\vspace{0.5cm}

\begin{center}
\begin{tabular}{lccccccccccc}
\hline
\hline
  &  &  &  &  &  &  & HML & ICC & ICC &  & All  \\[-2mm]
  & Rm &  & ICC & SMB & HML &  & + & + & + &  & four  \\[-2mm]
  &  &  &  &  &  &  & SMB & SMB & HML &  & factors \\ 
\cline{2-2}
\cline{4-6}
\cline{8-10}
\cline{12-12}
\multirow{2}{4cm}{Consumer Non Durables} & 75.9\% &  & \textbf{ 94.1\%} & 88.4\% & 79.7\% &  & 91.8\% & 94.1\% & \textbf{ 94.3\%} & \textbf{ } & 94.3\% \\[-2mm]
& {\tiny (70.9\%,80.5\%)} &&{\tiny (92.4\%,95.5\%)} &{\tiny (85.2\%,91.2\%)} &{\tiny (74.9\%,83.9\%)} &&{\tiny (89.5\%,93.9\%)} &{\tiny (92.5\%,95.5\%)} &{\tiny (92.7\%,95.6\%)} &&{\tiny (92.7\%,95.7\%)}\\[1mm]
\multirow{2}{4cm}{Consumer Durables} & 74.4\% &  & \textbf{ 92.3\%} & 87.9\% & 76.9\% &  & 90.2\% & \textbf{ 92.4\%} & 92.3\% &  & 92.4\% \\[-2mm]
& {\tiny (69.2\%,79.2\%)} &&{\tiny (90.2\%,94.2\%)} &{\tiny (84.8\%,91.1\%)} &{\tiny (72.2\%,81.9\%)} &&{\tiny (87.6\%,92.6\%)} &{\tiny (90.3\%,94.3\%)} &{\tiny (90.2\%,94.2\%)} &&{\tiny (90.4\%,94.3\%)}\\[1mm]
\multirow{2}{4cm}{Manufacturing} & 82.2\% &  & \textbf{ 96.7\%} & 92.0\% & 85.9\% &  & 95.4\% & 96.8\% & \textbf{ 97.0\%} & \textbf{ } & 97.1\% \\[-2mm]
& {\tiny (78.3\%,86.0\%)} &&{\tiny (95.7\%,97.6\%)} &{\tiny (89.9\%,93.9\%)} &{\tiny (82.5\%,88.9\%)} &&{\tiny (93.9\%,96.6\%)} &{\tiny (95.7\%,97.6\%)} &{\tiny (96.1\%,97.8\%)} &&{\tiny (96.2\%,97.9\%)}\\[1mm]
\multirow{2}{4cm}{Energy} & 58.3\% &  & \textbf{ 67.8\%} & 63.7\% & 63.4\% &  & 68.5\% & 68.1\% & \textbf{ 69.3\%} & \textbf{ } & 69.3\% \\[-2mm]
& {\tiny (51.7\%,64.5\%)} &&{\tiny (61.9\%,73.7\%)} &{\tiny (57.6\%,69.9\%)} &{\tiny (58.6\%,68.5\%)} &&{\tiny (63.0\%,74.1\%)} &{\tiny (62.3\%,74.0\%)} &{\tiny (64.0\%,74.7\%)} &&{\tiny (64.0\%,74.8\%)}\\[1mm]
\multirow{2}{4cm}{Business Equipment} & 74.5\% &  & \textbf{ 87.4\%} & 86.2\% & 74.8\% &  & 86.6\% & 88.0\% & \textbf{ 91.6\%} & \textbf{ } & 91.8\% \\[-2mm]
& {\tiny (68.7\%,79.9\%)} &&{\tiny (85.0\%,89.8\%)} &{\tiny (82.5\%,89.4\%)} &{\tiny (69.3\%,80.1\%)} &&{\tiny (83.2\%,89.6\%)} &{\tiny (85.9\%,90.4\%)} &{\tiny (90.1\%,93.0\%)} &&{\tiny (90.4\%,93.2\%)}\\[1mm]
\multirow{2}{4cm}{Telecom} & 62.7\% &  & \textbf{ 68.2\%} & 68.1\% & 63.9\% &  & 69.4\% & 68.6\% & \textbf{ 72.6\%} & \textbf{ } & 73.0\% \\[-2mm]
& {\tiny (55.4\%,69.2\%)} &&{\tiny (64.0\%,72.8\%)} &{\tiny (61.4\%,74.0\%)} &{\tiny (56.5\%,70.5\%)} &&{\tiny (63.4\%,75.1\%)} &{\tiny (64.2\%,74.2\%)} &{\tiny (69.0\%,77.0\%)} &&{\tiny (69.5\%,77.3\%)}\\[1mm]
\multirow{2}{4cm}{Shops} & 71.8\% &  & \textbf{ 90.1\%} & 86.7\% & 72.8\% &  & 87.6\% & 90.3\% & \textbf{ 90.4\%} & \textbf{ } & 90.5\% \\[-2mm]
& {\tiny (66.7\%,77.0\%)} &&{\tiny (86.2\%,93.1\%)} &{\tiny (82.8\%,90.5\%)} &{\tiny (67.7\%,78.2\%)} &&{\tiny (83.6\%,91.2\%)} &{\tiny (86.6\%,93.4\%)} &{\tiny (87.1\%,93.3\%)} &&{\tiny (87.1\%,93.4\%)}\\[1mm]
\multirow{2}{4cm}{Health} & 65.1\% &  & 74.5\% & \textbf{ 75.9\%} & 66.4\% &  & 77.4\% & 76.2\% & \textbf{ 80.5\%} & \textbf{ } & 80.5\% \\[-2mm]
& {\tiny (58.2\%,71.3\%)} &&{\tiny (69.5\%,79.0\%)} &{\tiny (72.0\%,79.7\%)} &{\tiny (60.0\%,72.5\%)} &&{\tiny (73.7\%,81.0\%)} &{\tiny (72.8\%,79.9\%)} &{\tiny (77.3\%,83.7\%)} &&{\tiny (77.4\%,83.8\%)}\\[1mm]
\multirow{2}{4cm}{Utilities} & 58.3\% &  &  60.8\% & 58.9\% & \textbf{65.9\%} &  & \textbf{ 66.5\%} & 61.7\% & 66.3\% &  & 66.5\% \\[-2mm]
& {\tiny (51.0\%,65.3\%)} &&{\tiny (52.8\%,68.8\%)} &{\tiny (51.6\%,66.8\%)} &{\tiny (58.2\%,72.8\%)} &&{\tiny (58.3\%,74.0\%)} &{\tiny (53.8\%,69.5\%)} &{\tiny (58.3\%,73.6\%)} &&{\tiny (58.5\%,74.1\%)}\\[1mm]
\multirow{2}{4cm}{Others} & 71.9\% &  & \textbf{ 92.8\%} & 83.6\% & 81.6\% &  & 92.7\% & 93.4\% & \textbf{ 95.2\%} & \textbf{ } & 95.2\% \\[-2mm]
& {\tiny (66.1\%,77.2\%)} &&{\tiny (90.5\%,94.8\%)} &{\tiny (79.3\%,87.5\%)} &{\tiny (77.5\%,85.3\%)} &&{\tiny (90.1\%,94.9\%)} &{\tiny (91.1\%,95.3\%)} &{\tiny (93.4\%,96.6\%)} &&{\tiny (93.5\%,96.6\%)}\\[1mm]
 &  &  &  &  &  &  &  &  &  &  &  \\[-2mm]
\multirow{2}{4cm}{Average} & 69.5\% &  & \textbf{ 82.4\%} & 79.1\% & 73.1\% &  & 82.6\% & 82.9\% & \textbf{ 84.9\%} & \textbf{ } & 85.0\% \\[-2mm]
& {\tiny (65.0\%,73.9\%)} &&{\tiny (79.9\%,85.1\%)} &{\tiny (76.0\%,82.3\%)} &{\tiny (69.0\%,77.1\%)} &&{\tiny (79.9\%,85.4\%)} &{\tiny (80.5\%,85.6\%)} &{\tiny (82.9\%,87.2\%)} &&{\tiny (83.0\%,87.3\%)}\\
\\
GRS&	2.53	&&	2.21&	2.69&	2.72	&&	2.61&	2.24&	2.70	&&	2.70\\
$p$-value&	0.01	&&	0.02&	0.00&	0.00	&&	0.00&	0.01&	0.00	&&	0.00\\
\hline
\hline
\end{tabular}
\end{center}
\end{table}

\end{landscape}

\clearpage

\clearpage

\clearpage

\end{document}